\documentclass[aps,prl,twocolumn,superscriptaddress,showpacs]{revtex4-1}
\usepackage{graphicx}%
\usepackage{color}

\def\CuCN{$\kappa$-(BEDT\--TTF)$_2$\-Cu$_2$(CN)$_3$}

\def\STF{$\kappa$-[(BEDT\--TTF)$_{1-x}$\-(BEDT\--STF)$_{x}$]$_2$\-Cu$_2$(CN)$_3$}
\def\stf{$\kappa$-STF$_x$}

\def\Cl{$\kappa$-(BEDT-TTF)$_2$Cu[N(CN)$_2$]Cl}

\def\HgCl{$\kappa$-(BEDT-TTF)$_2$Hg(SCN)$_2$Cl}

\begin{document}
\preprint{Percolative Mott transition}
\title{Low-Temperature Dielectric Anomalies at the Mott Insulator-Metal Transition}

\author{A. Pustogow}
\thanks{authors contributed equally}
\affiliation{1.~Physikalisches Institut, Universit\"{a}t Stuttgart, 70569 Stuttgart, Germany}
\affiliation{Department of Physics and Astronomy, UCLA, Los Angeles, California 90095, U.S.A.}
\author{R. R\"osslhuber}
\thanks{authors contributed equally}
\affiliation{1.~Physikalisches Institut, Universit\"{a}t Stuttgart, 70569 Stuttgart, Germany}
\author{Y. Tan}
\thanks{authors contributed equally}
\affiliation{National High Magnetic Field Laboratory, Florida State University, Tallahassee, U.S.A.}
\author{E. Uykur}
\affiliation{1.~Physikalisches Institut, Universit\"{a}t Stuttgart, 70569 Stuttgart, Germany}
\author{M. Wenzel}
\affiliation{1.~Physikalisches Institut, Universit\"{a}t Stuttgart, 70569 Stuttgart, Germany}
\author{A. B\"ohme}
\affiliation{1.~Physikalisches Institut, Universit\"{a}t Stuttgart, 70569 Stuttgart, Germany}
\author{A. L\"ohle}
\affiliation{1.~Physikalisches Institut, Universit\"{a}t Stuttgart, 70569 Stuttgart, Germany}
\author{R. H\"ubner}
\affiliation{1.~Physikalisches Institut, Universit\"{a}t Stuttgart, 70569 Stuttgart, Germany}
\affiliation{Institut f\"{u}r Klinische Radiologie und Nuklearmedizin, Universit\"{a}t Heidelberg, Germany}
\author{Y. Saito}
\affiliation{1.~Physikalisches Institut, Universit\"{a}t Stuttgart, 70569 Stuttgart, Germany}
\affiliation{Department of Physics, Hokkaido University, Sapporo, Japan}
\author{A. Kawamoto}
\affiliation{Department of Physics, Hokkaido University, Sapporo, Japan}
\author{J. A. Schlueter}
\affiliation{Material Science Division, Argonne National
Laboratory, Argonne, Illinois 60439-4831, U.S.A.}
\affiliation{Division of Materials Research, National Science Foundation, Alexandria, Virginia 22314, U.S.A.}
\author{V. Dobrosavljevi\'c}
\email{vlad@magnet.fsu.edu}
\affiliation{National High Magnetic Field Laboratory, Florida State University, Tallahassee, U.S.A.}
\author{M. Dressel}
\email{dressel@pi1.physik.uni-stuttgart.de}
\affiliation{1.~Physikalisches Institut, Universit\"{a}t Stuttgart, 70569 Stuttgart, Germany}
\date{\today}

\begin{abstract}
The correlation-driven Mott transition is commonly characterized by a drop in resistivity across the insulator-metal phase boundary; yet, the complex permittivity provides a deeper insight into the microscopic nature.
We investigate the frequency- and temperature-dependent dielectric response of the Mott insulator $\kappa$-(BEDT-TTF)$_{2}$\-Cu$_2$(CN)$_3$ when tuning from a quantum spin liquid into the Fermi-liquid state by applying external pressure and chemical substitution of the donor molecules.
At low temperatures the coexistence region at the first-order transition leads to a strong enhancement of the quasi-static dielectric constant $\epsilon_1$ when the effective correlations are tuned through the critical value. Several dynamical regimes are identified around the Mott point and vividly mapped through pronounced permittivity crossovers. All experimental trends are captured  by dynamical mean-field theory of the single-band Hubbard model supplemented by percolation theory.
\end{abstract}

\date{\today}

\pacs{
71.30.+h,  
75.25.Dk,
74.70.Kn,  
77.22.-d    
}

\maketitle

\date{\today}

The insulator-metal transition (IMT) remains the main unresolved basic science problem of condensed-matter physics. Especially intriguing are those IMTs not associated with static symmetry changes, where conventional paradigms for phase transitions provide little guidance. Early examples of such behavior are found in certain disorder-driven IMTs \cite{Shklovskii1984}. In recent years, IMTs with no symmetry breaking were also identified around the Mott transition \cite{Imada1998}, which bears close connection to exotic states of strongly-correlated electron matter such as superconductivity in the cuprates. From a theoretical point of view, the single-band Hubbard model is at present well understood \cite{Georges1996,Vollhardt2012}, and is found to be in excellent agreement with experiments \cite{Limelette2003,*Limelette2003a,Hansmann2013,
Kagawa2005,*Kagawa2004,*Furukawa2015}. While commonly concealed by antiferromagnetism, recent development in the field of organic quantum spin liquids (QSL) enabled us to study the low-temperature Mott IMT in absence of magnetic order~\cite{Kurosaki2005,*Shimizu2003,Shimizu2016,Itou2017,Li2019,Furukawa2018,Pustogow2018}, revealing finite-frequency precursors of the metal already on the insulating side~\cite{Pustogow2018}.

\begin{figure}[h]
\centering
\includegraphics[width=0.8\columnwidth]{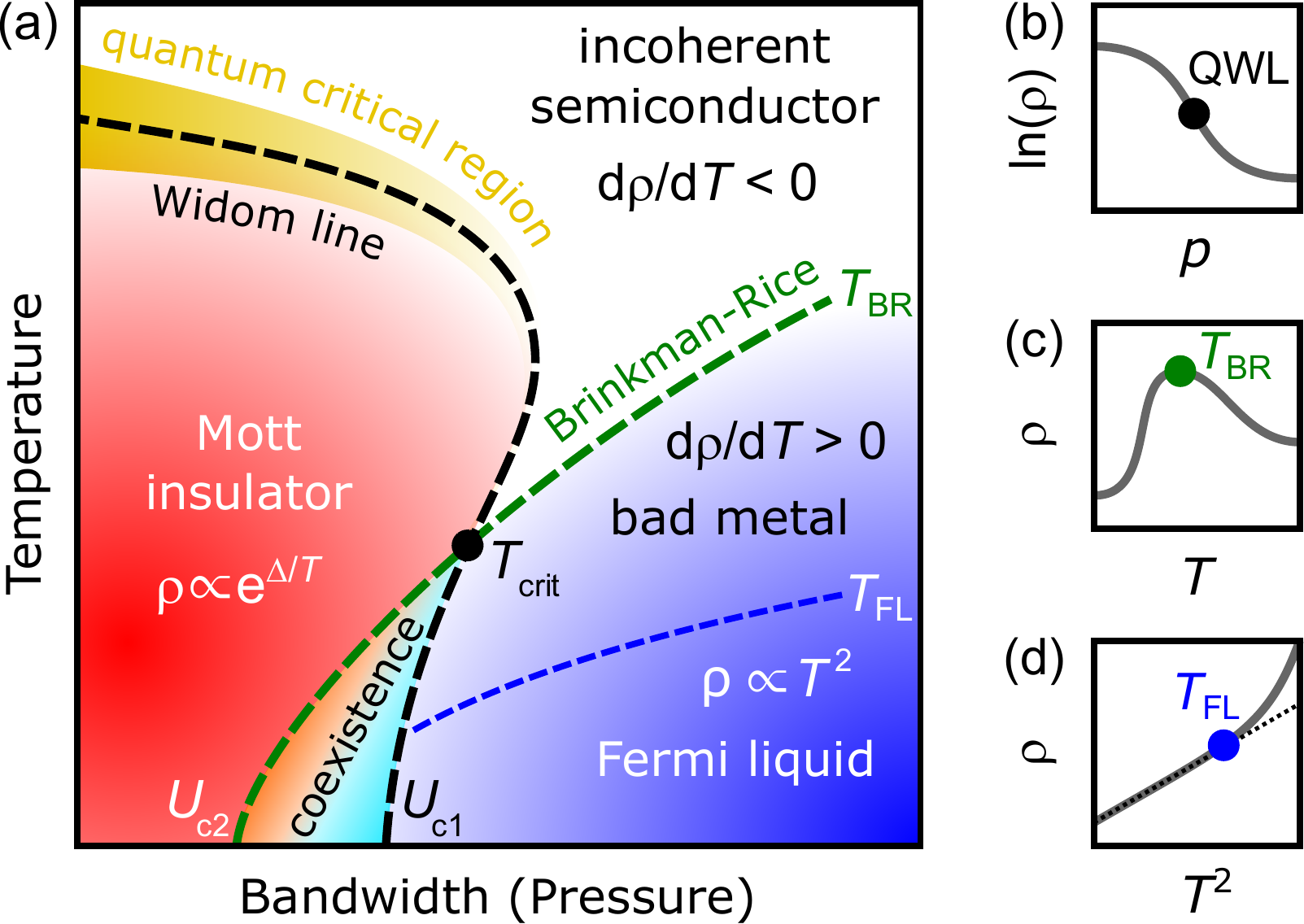}
\caption{(a) Tuning the bandwidth $W$, for instance by chemical or physical pressure, transforms a Mott insulator to a correlated metal. Dynamical mean-field theory predicts a first-order transition with phase coexistence up to the critical endpoint~\cite{Georges1996}, and a quantum-critical regime associated with the quantum Widom line (QWL) above $T_{\rm crit}$~\cite{Vucicevic2013}. The metallic state is confined by the Brinkman-Rice temperature $T_{\rm BR}$, the coherent Fermi-liquid regime by $T_{\rm FL}$. When interactions $U$ are comparable to $W$, and $ T \gg T_{\rm crit}$, semiconducting behavior prevails; neither a gap nor a quasiparticle peak are stabilized. (b-d) Resistivity signatures of the crossovers.}
\label{fig:phase-diagram}
\end{figure}

The Mott insulator and the correlated metal converge at the critical endpoint $T_{\rm crit}$ (Fig.~\ref{fig:phase-diagram}). The former is bounded by a quantum-critical region along the quantum Widom line (QWL)  \cite{Terletska2011,*Vucicevic2013,Furukawa2015,Pustogow2018,Dobrosavljevic1997}. On the metallic side, resistivity maxima at the “Brinkman-Rice” temperature $T_{\rm BR}$ signal the thermal destruction of resilient quasiparticles \cite{Radonjic2012,*Deng2013} and the crossover to semiconducting transport.
Below $T_{\rm crit}$, the IMT is of first order and comprises an insulator-metal coexistence regime \cite{Georges1996,Terletska2011,Vucicevic2013}. It is currently debated whether electrodynamics is dominated by closing of the Mott gap or by spatial inhomogeneity, fueled by recent low-temperature transport studies~\cite{Furukawa2018}.


%

Phase coexistence around the first-order line ($T<T_{\rm crit}$) emerges from bistability of the insulating and metallic phases between closing of the Mott gap at $U_{c1}$ and demise of the metal at $U_{c2}$ \cite{Georges1996}.
One generally expects hysteretic behaviors when tuning across a first-order transition. Seminal transport and susceptibility experiments indeed found a pronounced hysteresis in Mott insulators with a magnetically-ordered ground state \cite{Lefebvre2000,Limelette2003,*Limelette2003a}.
Unfortunately, analogue measurements with continuous pressure tuning are not feasible on QSL compounds, such as \CuCN, due to the low temperatures ($T<20$~K) and high pressures ($p>1$~kbar) required to cross the first-order IMT.

A more direct insight into the coexistence region was provided by spatially resolved optical spectroscopy \cite{Sasaki2004}. The most compelling results came from near-field optical experiments on vanadium oxides by Basov and collaborators \cite{Qazilbash2007,*Huffman2018,McLeod2016} where a spatial separation of metallic and insulating regions upon heating could be visualized,
in accord with x-ray studies \cite{Lupi2010,Hansmann2013}; the range with hysteresis in $\rho(T)$ coincides with the observed phase coexistence.
Although recent developments in cryogenic near-field instrumentation are rather promising~\cite{McLeod2016,Post2018,*Pustogow2018s}, they fall short of covering the regime $T<T_{\rm crit} \approx 15$~K required here and do not allow for pressure tuning.
For this reason, we suggest dielectric spectroscopy as novel bulk-sensitive method in order to reveal the coexistence regime, distinguish the individual phases and obtain a deeper understanding of the dynamics around the IMT.
The complex conductivity $\sigma_1+{\rm i}\sigma_2$ not only reveals the closing of the Mott gap but yields insight into the growth of metallic regions and the formation of quasiparticles as correlation effects decrease.

 In this Letter we tackle the fundamental question whether the electrodynamic response around the Mott IMT is overwhelmed by the gradual decrease of the Mott-Hubbard gap within a homogeneous insulating phase, or whether the effects of phase coexistence dominate. Furthermore, is it possible to distinguish on the metallic side between the coherent (quasiparticle) low-$T$ regime and incoherent transport at high-$T$? To answer these questions, we present temperature- and frequency-dependent dielectric measurements on a genuine Mott compound that is bandwidth-tuned across its first-order IMT. In addition to hydrostatic pressure we developed a novel approach of chemically substituting the organic donor molecules. The experimental findings are complemented by dynamical mean-field theory (DMFT) calculations, incorporating spatial inhomogeneities in a hybrid approach. We conclude that electronic phase segregation plays a crucial role, leading to percolative phenomena due to the separation of insulating and metallic regions, also allowing clear and precise mapping of different dynamical regimes around the IMT.

\begin{figure}
\centering
\includegraphics[width=1\columnwidth]{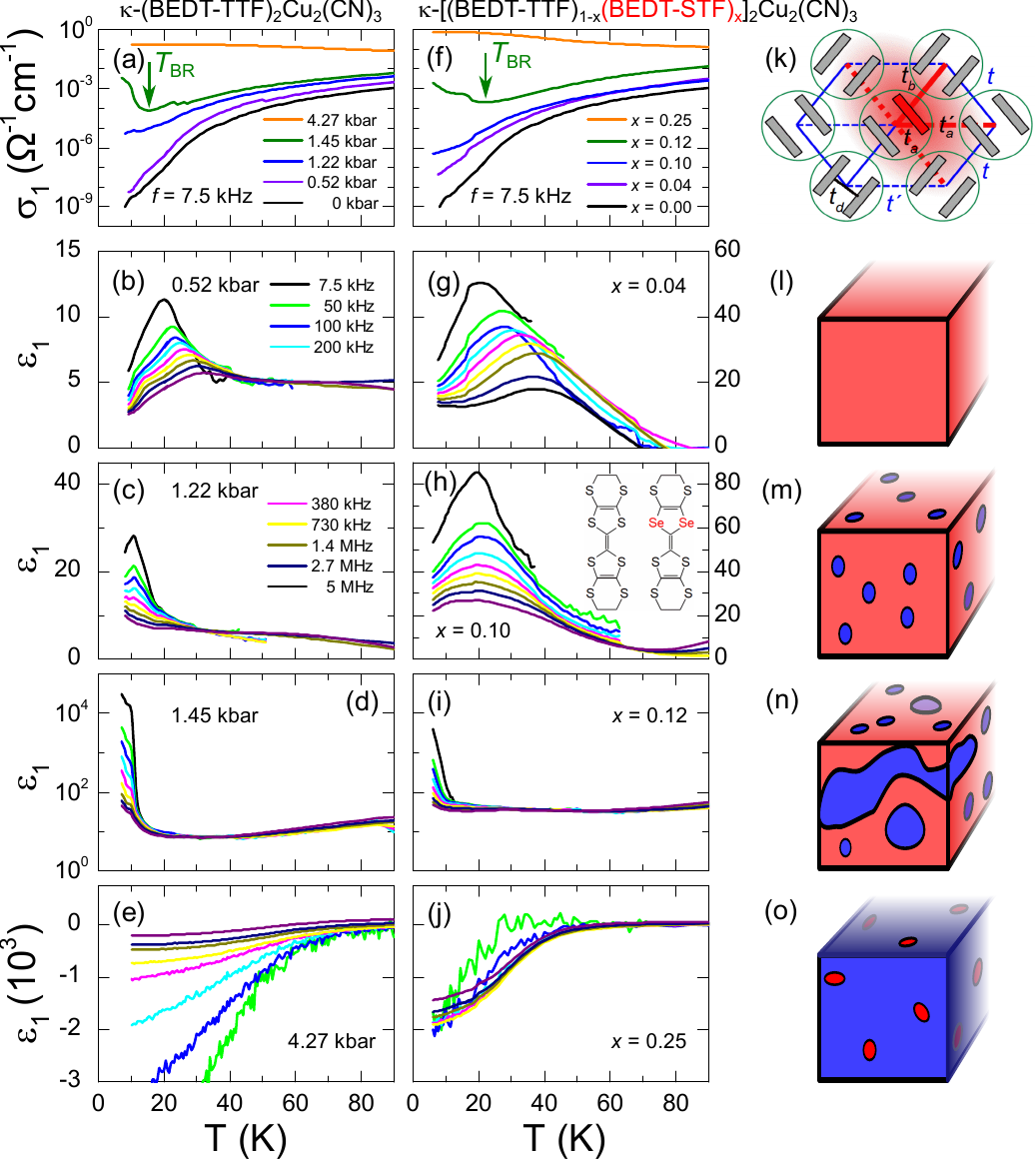}
\caption{The dielectric conductivity and permittivity of \CuCN\ were measured as a function of temperature and frequency for various applied (a-e) pressures and (f-j) chemical substitutions [introduction of Se-containing BEDT-STF molecules illustrated in (h) and (k)] that drive the system across the Mott transition. (a,f)~Starting from the insulator, $\sigma_1(T)$ grows with increasing $p$ or $x$; a metallic phase is stabilized below $T_{\rm BR}$, in accord with dc transport~\cite{Kurosaki2005,Furukawa2015,Furukawa2018}.
(b,c,g,h) In the Mott-insulating state $\epsilon_1(T)$ exhibits relaxor-ferroelectric behavior similar to the parent compound~\cite{Abdel-Jawad2010,Pinteric2014}. Extrinsic high-temperature contributions are subtracted.
(d,i) The strong enhancement of $\epsilon_1(T)$ at the transition is a hallmark of a percolative IMT, as sketched in (l-o).
(e,j) When screening becomes dominant in the metal,
$\epsilon_1$ turns negative; $\sigma_1$ exhibits Fermi-liquid behavior below $T_{\rm FL}$. }
\label{fig:sigma-eps_T}
\end{figure}

We have chosen \CuCN\ single crystals for our investigations
because this paradigmatic QSL candidate is well characterized by electric, optical and magnetic measurements \cite{Kurosaki2005,*Shimizu2003,Kezsmarki2006,Kanoda2011,*Zhou2017,Furukawa2018,
Pustogow2018,Culo2019}.
Although the dimerized charge-transfer salt possesses a half-filled conduction band, strong electronic interaction $U \approx 250$~meV stabilizes a Mott-insulating state below the QWL ($T_{\rm QWL}\approx 185$~K at ambient pressure~\cite{Pustogow2018,Furukawa2015}).
The effective correlation strength $U/W$ can be reduced by increasing the bandwidth $W$; for pressure $p > 1.4$~kbar the metallic state is reached at low temperatures  \footnote{The superconducting state at $T\approx 4$~K \cite{Kurosaki2005,Furukawa2018} is  below the temperature accessible to us here.}, with $T_{\rm crit} \approx 15$-20~K.
In addition, we exploited a novel route of partially replacing the S atoms of the donor molecules by Se, where more extended orbitals lead to larger bandwidth [see sketches in Fig.~\ref{fig:sigma-eps_T}(h,k)]. The substitutional series \STF\ ($0\leq x \leq 1$, abbreviated \stf) spans the interval ranging from a Mott insulator to a Fermi-liquid metal~\footnote{BEDT-TTF stands for bis-ethylene-dithio-tetra\-thia\-fulvalene. Substituting two of the inner sulfur atoms by selenium leads to bis-ethylene\-dithio-di\-selenium-di\-thia\-fulvalene, abbreviated BEDT-STF \cite{Saito2019}.}.
Details on the sample characterization and experimental methods are given in Refs.~\onlinecite{Rosslhuber2019,Saito2018,Saito2019}.
Here we focus on the out-of-plane dielectric response measured from $f=7$~kHz to 5~MHz down to $T=5$~K. Both physical pressure and STF-substitution allow us to monitor the permittivity while shifting the system across the first-order IMT.


Fig.~\ref{fig:sigma-eps_T} displays the temperature-dependent conductivity and  permittivity data
of \CuCN\ when  $p$ rises (a-e) and $x$ increases in \stf\ (f-j).
The insulating state ($p<1.4$~kbar, $x<0.1$), characterized by ${\rm d}\sigma_1/{\rm d}T>0$, generally features small positive $\epsilon_1\approx 10$.
The relaxor-like response previously observed in the parent compound below 50~K has been subject of debate \cite{Abdel-Jawad2010,Pinteric2014}.
The metallic state ($p>3$~kbar, $x>0.2$) is defined by ${\rm d}\sigma_1/{\rm d}T<0$ and, concomitantly, $\epsilon_1<0$ that becomes very large at low $T$ as itinerant electrons increasingly screen~\footnote{Comparison of the results in Fig.~\ref{fig:sigma-eps_T}(j) with optical data measured on the same substitution yields fair agreement of the metallic values $\epsilon_1 < 0$~\cite{Saito2019}. Technical details of the dielectric experiments can be found in Ref.~\cite{Rosslhuber2019}.}.
This onset of metallic transport identifies $T_{\rm BR}$~\cite{Radonjic2012}; while thermal fluctuations prevail at higher $T$, the quasiparticle bandwidth is the dominant energy scale for $T<T_{\rm BR}$. Below $T_{\rm FL}$ the resistivity $\rho(T)\propto T^2$ indicates the Fermi-liquid state.

Right at the first-order IMT, however, the dielectric behavior appears rather surprising. When approaching the low-temperature phase boundary, $\epsilon_1$ rapidly increases by several orders of magnitude. This colossal permittivity enhancement
is more pronounced in the quasi-static limit, $\epsilon_1\approx 10^5$ at $f=7.5$~kHz, and the peak value approximately follows  a $f^{-1.5}$ dependence.
The overall range in $T$ and $p$/$x$ of the divergency is robust and does not depend on the probing frequency; detailed analysis on the dynamic properties is given in \cite{Rosslhuber2019,Saito2019}.

\begin{figure}
\centering
\includegraphics[width=1\columnwidth]{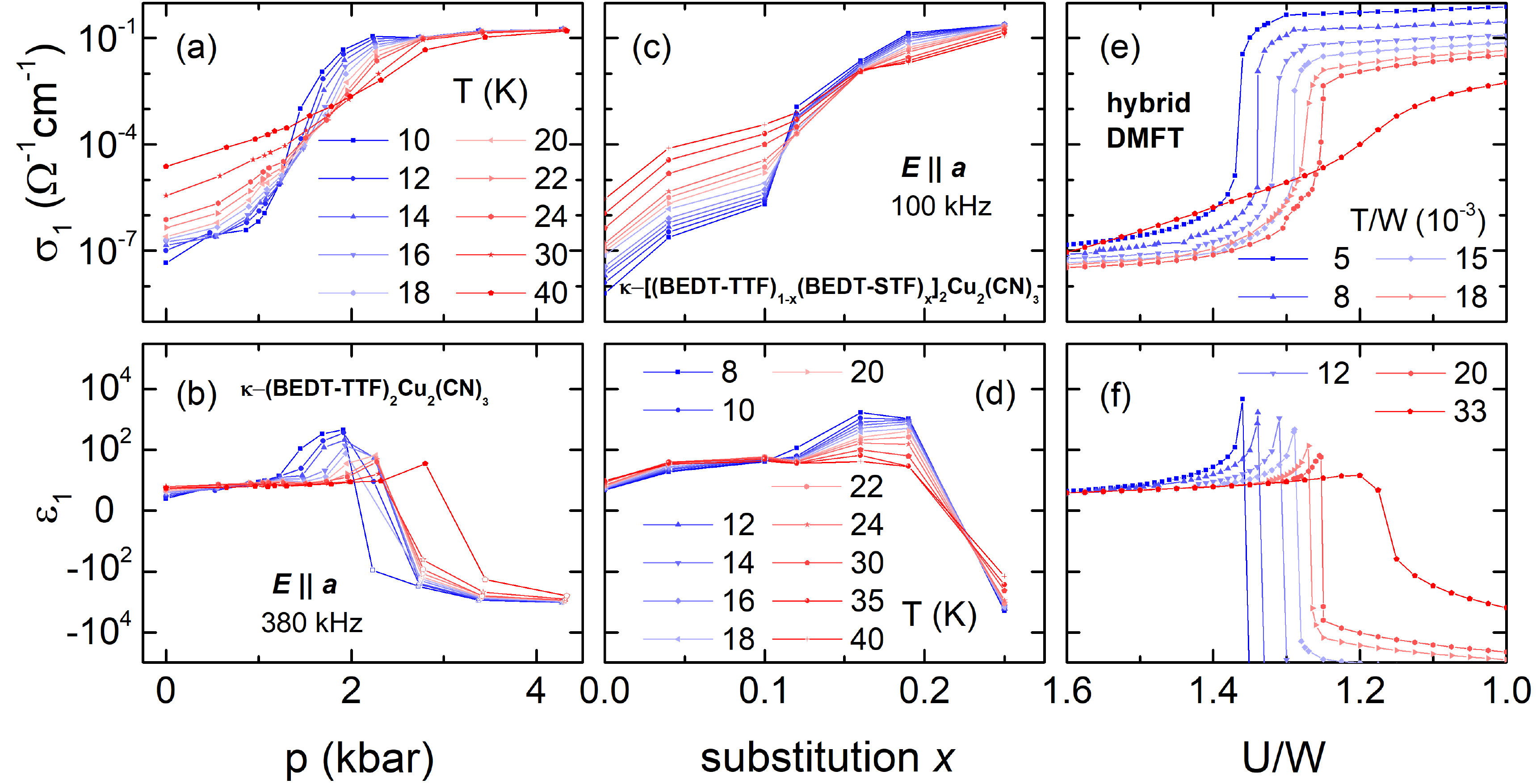}
\caption{(a) The Mott IMT of $\kappa$-(BEDT-TTF)$_2$Cu$_2$(CN)$_3$ appears as a rapid increase of $\sigma_1(p)$ that smoothens at higher $T$; above $T_{\rm crit}$ a gradual crossover remains. (b) $\epsilon_1(p)$ exhibits a sharp peak below $T_{\rm crit}$. The results at $f=380$~kHz are plotted on a logarithmic scale. (c,d) Similar behavior is observed for chemical BEDT-STF substitution. (e,f) Fixed-temperature line cuts of our hybrid DMFT simulations (see text) as a function of correlation strength $U/W$ and $T/W$~\cite{Vucicevic2013,Pustogow2018} resemble the experimental situation in minute detail, including the shift of the IMT with $T$. The lack of saturation of $\sigma_1(T\rightarrow 0)$ seen in DMFT modeling reflects the neglect of elastic (impurity) scattering in the metal (outside the coexistence region).}
\label{fig:sigma-eps_p_x_DMFT}
\end{figure}

In Fig.~\ref{fig:sigma-eps_p_x_DMFT}(a,b) the pressure evolution of $\sigma_1$ and $\epsilon_1$ is plotted for fixed $T$. At $T=10$~K, $\sigma_1(p)$ rises by six orders of magnitude in the narrow range of 1~kbar and $\Delta x = 0.1$. This behavior flattens to a gradual transition above 20~K, associated with the quantum-critical crossover at the QWL. The inflection point shifts to higher $p$, in accord with the positive slope of the phase boundary~\cite{Pustogow2018} associated with the rising onset of metallicity at $T_{\rm BR}$. The \stf\ series exhibits similar behavior [Fig.~\ref{fig:sigma-eps_p_x_DMFT}(c,d)]: around the critical concentration of $x\approx 0.12$ a drastic increase in $\sigma_1$ is observed at low $T$ that smears out as $T$ rises. The maximum in $\epsilon_1(x)$ is reached for $x=0.16$ but broadens rapidly upon heating. 

\begin{figure*}
\centering
\includegraphics[width=0.9\textwidth]{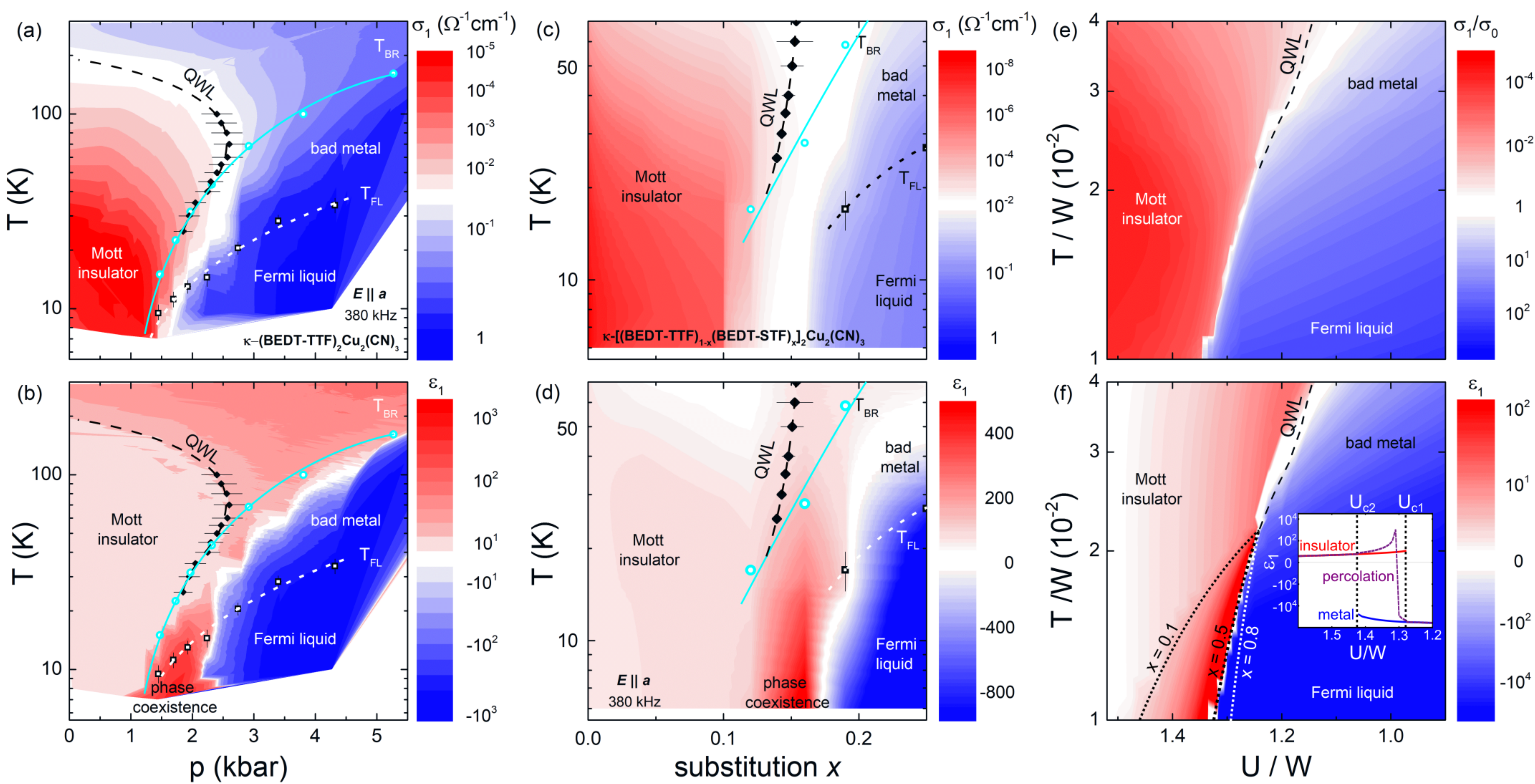}
\caption{Phase diagram of $\kappa$-(BEDT-TTF)$_2$Cu$_2$(CN)$_3$ when tuned through the Mott IMT by physical pressure (a,b) or chemical substitution (c,d), compared with hybrid DMFT calculations as a function of correlation strength $U/W$ (e,f); the respective color codes are given to the right. In all cases, the colossal enhancement of $\epsilon_1$ reveals a sharply defined insulator-metal phase-coexistence region around the first-order IMT, which is not seen in $\sigma_1$.
The permittivity clearly distinguishes the Mott-insulating and metallic states via small positive and large negative values, which line up perfectly with the quantum Widom line (QWL) and the Brinkman-Rice temperature $T_{\rm BR}$ determined from $\sigma_1(T,p,x)$, respectively. Inset of (f): metallic and insulating DMFT solutions are contrasted to the mixed phase (metallic fractions indicated in main panel) that reproduces the strong peak in $\epsilon_1$.
}
\label{fig:contour-plots}
\end{figure*}

The Mott IMT is based on the idea that a reduction of electronic correlations, i.e.\ rise of $W/U$, gradually closes the Mott-Hubbard gap: a coherent charge response develops, causing a finite metallic conductivity. Pressure-dependent
optical studies on several organic Mott insulators actually observe this behavior \cite{Faltermeier2007,*Merino2008,*Dumm2009,Li2019}.
It was pointed out \cite{Aebischer2001} that even in the case of certain second-order phase transitions, a continuously vanishing charge gap might induce an enhancement in $\epsilon_1$, perhaps leading to a  divergence at low $T$. Probing the optical response at THz frequencies is actually a convenient method to monitor the gap contribution to the permittivity. From $p$ and $T$ sweeps across the Mott IMT of very different materials an increase by a factor of 10 is consistently reported \cite{Qazilbash2007,Li2019}; in the case of \stf\ we find it even smaller \cite{Saito2019}. Hence, the dielectric catastrophe of $\epsilon_1 \approx 10^5$ observed in our pressure and substitution-dependent dielectric experiments evidences an additional effect.

Treating the fully-frustrated model at half filling, even at $T=0$ DMFT finds metallic and insulating solutions coexisting over an appreciable range of $U/W$
\cite{Georges1996}. This may result in a spatial segregation of these distinct electronic phases. Such a picture resembles composite materials, such as microemulsions, composites or percolating metal films \cite{vanDijk1986,Clarkson1988a,*Clarkson1988b,
Pecharroman2000,Nan2010,
Hovel2010}.
Classical percolation is not a thermodynamic phase transition, but a statistical problem that has been studied theoretically for decades by analytical and numerical methods in all details; one of the key predictions is the divergency of $\epsilon_1$ when approaching the transition from either side, with characteristic scaling \cite{Dubrov1976,Efros1976,Bergman1977,*Bergman1978}.
Over a large parameter range the dielectric properties are well described by the Maxwell-Garnett or Bruggeman effective medium approaches \cite{Choy2015}.

To further substantiate this physical picture in quantitative details, we carried out theoretical modeling of the systems under study. We calculated $\epsilon_1+{\rm i}\epsilon_2$ using DMFT for a single-band Hubbard model, and obtained the respective responses for both the insulating and the metallic phase around the Mott point~\cite{Vucicevic2013}.
The DMFT phase diagram (Fig.~\ref{fig:phase-diagram}) also features an intermediate coexistence region below $ k_{B} T_{\rm crit} \approx 0.02 W$. In accord with the analysis of our experimental results, we assumed a smoothly varying  metallic fraction $x$ within the phase coexistence region. To obtain the total dielectric response, we solved an appropriate electrical-network model representing such spatial inhomogeneity utilizing the standard effective-medium approximation (for details, see Ref.~\onlinecite{Rosslhuber2019}).

Our hybrid DMFT simulation yields excellent agreement with experiment -- both pressure-tuning and chemical substitution -- as illustrated in Fig.~\ref{fig:sigma-eps_p_x_DMFT}(e,f) and in the false-color plots in Fig.~\ref{fig:contour-plots}.
We find that the colossal peak in $\epsilon_1$ is confined to the spatially inhomogeneous coexistence regime, exactly as observed in experiments. As correlation effects diminish, the dynamical conductivity (upper panels) increases from the Mott insulator to the Fermi liquid. The step in $\sigma_1$ and drop in $\epsilon_1$ appear abruptly in the model but more smoothly in experiment, most likely due to inhomogeneities which broaden the coexistence regime by providing nucleation seeds for the incipient phase.
Note, the percolative transition region narrows for $T\rightarrow T_{\rm crit}$ and vanishes above that; the metallic fraction of the simulation is indicated in Fig.~\ref{fig:contour-plots}(f).

The inset of panel (f) clearly demonstrates that the colossal permittivity enhancement appears exclusively for a percolating mixture of metallic and insulating regions, and not for the pure phases. These results render the gap closing irrelevant for the electrodynamics at the low-temperature Mott IMT. While the transition is of first-order at half filling, doping \cite{Hebert2015} or disorder \cite{Gati2018,Urai2019} effectively move $T_{\rm crit}\rightarrow 0$, eventually turning it into a true quantum-critical point.
We further point out that the discussed mechanism of a percolative enhancement of $\epsilon_1$ may also apply to related organic compounds subject to first-order transitions. Similar dielectric anomalies in \HgCl\ and \Cl\ were previously assigned to ferroelectricity~\cite{Gati2018a} and multiferroicity~\cite{Lunkenheimer2012}, respectively. The former exhibits a peak in $\epsilon_1$ right at its charge-order IMT, where phase coexistence is evident~\cite{Gati2018a,Hassan2019}. The latter is located extremely close to the Mott IMT~\cite{Limelette2003a}, so the coexistence region is likely entered already at ambient pressure~\footnote{While $\epsilon_1$ initially increases upon cooling in \Cl, it peaks at the antiferromagnetic transition and reduces at lower $T$. This could be a consequence of the metallic fraction first increasing as the insulator-metal phase boundary approaches the ambient-pressure position, but then reducing below $T_{\rm N}$ because of the negative slope of the boundary between antiferromagnet and metal which moves the IMT further away from $p=0$.}.

While previous transport and optical studies~\cite{Furukawa2015,Pustogow2018} already provided hints favoring the DMFT scenario, they do not map out the predicted dynamical regimes, especially regarding a well-defined coexistence region at $T < T_{\rm crit}$. Our new dielectric data, however, reveal all phases in vivid detail and in remarkable agreement with the respective crossover lines obtained from dc transport. Indeed, we recognize the gapped Mott insulator by essentially constant $\epsilon_1$ (light red) bounded precisely by the QWL~\cite{Vucicevic2013}, while also below $T_{\rm BR}$ \cite{Radonjic2012,*Deng2013} the response clearly follows the dielectric behavior expected for a metal ($\epsilon_1 < 0$, blue). Most remarkably, these two crossover lines converge towards $T_{\rm crit}$, which marks the onset of the coexistence region, just as anticipated from Fig.~\ref{fig:phase-diagram}.  The emergence of phase segregation is evidenced by the huge peak of $\epsilon_1$ in excellent agreement with our current DMFT-based modeling. The sharply defined boundaries of this dielectric anomaly imply that the corresponding inhomogeneities are  {\em the consequence and not the cause} of phase separation, the latter resulting from strong correlation effects inherent to Mottness. Our findings leave little room for doubt that the DMFT scenario offers a rather accurate picture of the Mott IMT, in contrast to other theoretical viewpoints \cite{Senthil2008} which focus on the spin degrees of freedom in the QSL. This also confirms recent experimental and theoretical results~\cite{Lee2016,Pustogow2018spinons} suggesting that such gapless spin excitations, while dominant deep within the low-temperature Mott-insulating phase, are quickly damped away by the onset of charge fluctuations close to the IMT.

We also demonstrated that the novel chemical method of partially substituting the organic donor molecules of the fully-frustrated Mott insulator \CuCN\ by BEDT-STF yields similar bandwidth-tuning like physical pressure. By comparing the boundaries of the Mott state and the correlated metal (QWL, $T_{\rm BR}$, $T_{\rm FL}$) we find that 1~kbar is equivalent to $\Delta x \approx 0.06$. The pronounced divergency in $\epsilon_1$ evidences a spatial coexistence of metallic and insulating electronic phases around the first-order IMT that can be circumstantially described by percolation theory. Our results yield that the Mott gap has a minor effect on the dielectric properties while the effects of phase coexistence dominate.

\acknowledgments
We appreciate discussions with S. Brown, B. Gompf and I. Voloshenko. We acknowledge support by the
DFG 
via DR228/52-1. A.P. acknowledges support by the Alexander von Humboldt Foundation through the Feodor Lynen Fellowship. Work in Florida was supported by the NSF Grant No. 1822258, and the National High Magnetic Field Laboratory through the NSF Cooperative Agreement No. 1157490 and the State of Florida. E.U. receives support of the European Social Fund and the Ministry of Science Research and the Arts of Baden-W\"urttemberg. J.A.S. acknowledges support from the Independent Research/Development program while serving at the National Science Foundation. A.P., R.R. and Y.T. contributed equally to this work.


\begin{thebibliography}{67}%
\makeatletter
\providecommand \@ifxundefined [1]{%
 \@ifx{#1\undefined}
}%
\providecommand \@ifnum [1]{%
 \ifnum #1\expandafter \@firstoftwo
 \else \expandafter \@secondoftwo
 \fi
}%
\providecommand \@ifx [1]{%
 \ifx #1\expandafter \@firstoftwo
 \else \expandafter \@secondoftwo
 \fi
}%
\providecommand \natexlab [1]{#1}%
\providecommand \enquote  [1]{``#1''}%
\providecommand \bibnamefont  [1]{#1}%
\providecommand \bibfnamefont [1]{#1}%
\providecommand \citenamefont [1]{#1}%
\providecommand \href@noop [0]{\@secondoftwo}%
\providecommand \href [0]{\begingroup \@sanitize@url \@href}%
\providecommand \@href[1]{\@@startlink{#1}\@@href}%
\providecommand \@@href[1]{\endgroup#1\@@endlink}%
\providecommand \@sanitize@url [0]{\catcode `\\12\catcode `\$12\catcode
  `\&12\catcode `\#12\catcode `\^12\catcode `\_12\catcode `\%12\relax}%
\providecommand \@@startlink[1]{}%
\providecommand \@@endlink[0]{}%
\providecommand \url  [0]{\begingroup\@sanitize@url \@url }%
\providecommand \@url [1]{\endgroup\@href {#1}{\urlprefix }}%
\providecommand \urlprefix  [0]{URL }%
\providecommand \Eprint [0]{\href }%
\providecommand \doibase [0]{http://dx.doi.org/}%
\providecommand \selectlanguage [0]{\@gobble}%
\providecommand \bibinfo  [0]{\@secondoftwo}%
\providecommand \bibfield  [0]{\@secondoftwo}%
\providecommand \translation [1]{[#1]}%
\providecommand \BibitemOpen [0]{}%
\providecommand \bibitemStop [0]{}%
\providecommand \bibitemNoStop [0]{.\EOS\space}%
\providecommand \EOS [0]{\spacefactor3000\relax}%
\providecommand \BibitemShut  [1]{\csname bibitem#1\endcsname}%
\let\auto@bib@innerbib\@empty
\bibitem [{\citenamefont {Shklovskii}\ and\ \citenamefont
  {Efros}(1984)}]{Shklovskii1984}%
  \BibitemOpen
  \bibfield  {author} {\bibinfo {author} {\bibfnamefont {B.~I.}\ \bibnamefont
  {Shklovskii}}\ and\ \bibinfo {author} {\bibfnamefont {A.~L.}\ \bibnamefont
  {Efros}},\ }\href@noop {} {\emph {\bibinfo {title} {Electronic Properties of
  Doped Semiconductors}}}\ (\bibinfo  {publisher} {Springer-Verlag},\ \bibinfo
  {address} {Berlin},\ \bibinfo {year} {1984})\BibitemShut {NoStop}%
\bibitem [{\citenamefont {Imada}\ \emph {et~al.}(1998)\citenamefont {Imada},
  \citenamefont {Fujimori},\ and\ \citenamefont {Tokura}}]{Imada1998}%
  \BibitemOpen
  \bibfield  {author} {\bibinfo {author} {\bibfnamefont {M.}~\bibnamefont
  {Imada}}, \bibinfo {author} {\bibfnamefont {A.}~\bibnamefont {Fujimori}}, \
  and\ \bibinfo {author} {\bibfnamefont {Y.}~\bibnamefont {Tokura}},\ }\href
  {https://link.aps.org/doi/10.1103/RevModPhys.70.1039} {\bibfield  {journal}
  {\bibinfo  {journal} {Rev. Mod. Phys.}\ }\textbf {\bibinfo {volume} {70}},\
  \bibinfo {pages} {1039} (\bibinfo {year} {1998})}\BibitemShut {NoStop}%
\bibitem [{\citenamefont {Georges}\ \emph {et~al.}(1996)\citenamefont
  {Georges}, \citenamefont {Kotliar}, \citenamefont {Krauth},\ and\
  \citenamefont {Rozenberg}}]{Georges1996}%
  \BibitemOpen
  \bibfield  {author} {\bibinfo {author} {\bibfnamefont {A.}~\bibnamefont
  {Georges}}, \bibinfo {author} {\bibfnamefont {G.}~\bibnamefont {Kotliar}},
  \bibinfo {author} {\bibfnamefont {W.}~\bibnamefont {Krauth}}, \ and\ \bibinfo
  {author} {\bibfnamefont {M.~J.}\ \bibnamefont {Rozenberg}},\ }\href
  {https://link.aps.org/doi/10.1103/RevModPhys.68.13} {\bibfield  {journal}
  {\bibinfo  {journal} {Rev. Mod. Phys.}\ }\textbf {\bibinfo {volume} {68}},\
  \bibinfo {pages} {13} (\bibinfo {year} {1996})}\BibitemShut {NoStop}%
\bibitem [{\citenamefont {Vollhardt}(2012)}]{Vollhardt2012}%
  \BibitemOpen
  \bibfield  {author} {\bibinfo {author} {\bibfnamefont {D.}~\bibnamefont
  {Vollhardt}},\ }\href {\doibase 10.1002/andp.201100250} {\bibfield  {journal}
  {\bibinfo  {journal} {Ann. Phys. (Berl.)}\ }\textbf {\bibinfo {volume}
  {524}},\ \bibinfo {pages} {1} (\bibinfo {year} {2012})}\BibitemShut {NoStop}%
\bibitem [{\citenamefont {Limelette}\ \emph
  {et~al.}(2003{\natexlab{a}})\citenamefont {Limelette}, \citenamefont
  {Georges}, \citenamefont {J{\'{e}}rome}, \citenamefont {Wzietek},
  \citenamefont {Metcalf},\ and\ \citenamefont {Honig}}]{Limelette2003}%
  \BibitemOpen
  \bibfield  {author} {\bibinfo {author} {\bibfnamefont {P.}~\bibnamefont
  {Limelette}}, \bibinfo {author} {\bibfnamefont {A.}~\bibnamefont {Georges}},
  \bibinfo {author} {\bibfnamefont {D.}~\bibnamefont {J{\'{e}}rome}}, \bibinfo
  {author} {\bibfnamefont {P.}~\bibnamefont {Wzietek}}, \bibinfo {author}
  {\bibfnamefont {P.}~\bibnamefont {Metcalf}}, \ and\ \bibinfo {author}
  {\bibfnamefont {J.~M.}\ \bibnamefont {Honig}},\ }\href
  {http://science.sciencemag.org/content/302/5642/89.abstract} {\bibfield
  {journal} {\bibinfo  {journal} {Science}\ }\textbf {\bibinfo {volume}
  {302}},\ \bibinfo {pages} {89 } (\bibinfo {year}
  {2003}{\natexlab{a}})}\BibitemShut {NoStop}%
\bibitem [{\citenamefont {Limelette}\ \emph
  {et~al.}(2003{\natexlab{b}})\citenamefont {Limelette}, \citenamefont
  {Wzietek}, \citenamefont {Florens}, \citenamefont {Georges}, \citenamefont
  {Costi}, \citenamefont {Pasquier}, \citenamefont {J{\'{e}}rome},
  \citenamefont {M{\'{e}}zi{\`{e}}re},\ and\ \citenamefont
  {Batail}}]{Limelette2003a}%
  \BibitemOpen
  \bibfield  {author} {\bibinfo {author} {\bibfnamefont {P.}~\bibnamefont
  {Limelette}}, \bibinfo {author} {\bibfnamefont {P.}~\bibnamefont {Wzietek}},
  \bibinfo {author} {\bibfnamefont {S.}~\bibnamefont {Florens}}, \bibinfo
  {author} {\bibfnamefont {A.}~\bibnamefont {Georges}}, \bibinfo {author}
  {\bibfnamefont {T.~A.}\ \bibnamefont {Costi}}, \bibinfo {author}
  {\bibfnamefont {C.}~\bibnamefont {Pasquier}}, \bibinfo {author}
  {\bibfnamefont {D.}~\bibnamefont {J{\'{e}}rome}}, \bibinfo {author}
  {\bibfnamefont {C.}~\bibnamefont {M{\'{e}}zi{\`{e}}re}}, \ and\ \bibinfo
  {author} {\bibfnamefont {P.}~\bibnamefont {Batail}},\ }\href
  {https://link.aps.org/doi/10.1103/PhysRevLett.91.016401} {\bibfield
  {journal} {\bibinfo  {journal} {Phys. Rev. Lett.}\ }\textbf {\bibinfo
  {volume} {91}},\ \bibinfo {pages} {16401} (\bibinfo {year}
  {2003}{\natexlab{b}})}\BibitemShut {NoStop}%
\bibitem [{\citenamefont {Hansmann}\ \emph {et~al.}(2013)\citenamefont
  {Hansmann}, \citenamefont {Toschi}, \citenamefont {Sangiovanni},
  \citenamefont {Saha-Dasgupta}, \citenamefont {Lupi}, \citenamefont {Marsi},\
  and\ \citenamefont {Held}}]{Hansmann2013}%
  \BibitemOpen
  \bibfield  {author} {\bibinfo {author} {\bibfnamefont {P.}~\bibnamefont
  {Hansmann}}, \bibinfo {author} {\bibfnamefont {A.}~\bibnamefont {Toschi}},
  \bibinfo {author} {\bibfnamefont {G.}~\bibnamefont {Sangiovanni}}, \bibinfo
  {author} {\bibfnamefont {T.}~\bibnamefont {Saha-Dasgupta}}, \bibinfo {author}
  {\bibfnamefont {S.}~\bibnamefont {Lupi}}, \bibinfo {author} {\bibfnamefont
  {M.}~\bibnamefont {Marsi}}, \ and\ \bibinfo {author} {\bibfnamefont
  {K.}~\bibnamefont {Held}},\ }\href {\doibase 10.1002/pssb.201248476}
  {\bibfield  {journal} {\bibinfo  {journal} {Phys. Status Solidi B}\ }\textbf
  {\bibinfo {volume} {250}},\ \bibinfo {pages} {1251} (\bibinfo {year}
  {2013})}\BibitemShut {NoStop}%
\bibitem [{\citenamefont {Kagawa}\ \emph {et~al.}(2005)\citenamefont {Kagawa},
  \citenamefont {Miyagawa},\ and\ \citenamefont {Kanoda}}]{Kagawa2005}%
  \BibitemOpen
  \bibfield  {author} {\bibinfo {author} {\bibfnamefont {F.}~\bibnamefont
  {Kagawa}}, \bibinfo {author} {\bibfnamefont {K.}~\bibnamefont {Miyagawa}}, \
  and\ \bibinfo {author} {\bibfnamefont {K.}~\bibnamefont {Kanoda}},\
  }\href@noop {} {\bibfield  {journal} {\bibinfo  {journal} {Nature}\ }\textbf
  {\bibinfo {volume} {436}},\ \bibinfo {pages} {534} (\bibinfo {year}
  {2005})}\BibitemShut {NoStop}%
\bibitem [{\citenamefont {Kagawa}\ \emph {et~al.}(2004)\citenamefont {Kagawa},
  \citenamefont {Itou}, \citenamefont {Miyagawa},\ and\ \citenamefont
  {Kanoda}}]{Kagawa2004}%
  \BibitemOpen
  \bibfield  {author} {\bibinfo {author} {\bibfnamefont {F.}~\bibnamefont
  {Kagawa}}, \bibinfo {author} {\bibfnamefont {T.}~\bibnamefont {Itou}},
  \bibinfo {author} {\bibfnamefont {K.}~\bibnamefont {Miyagawa}}, \ and\
  \bibinfo {author} {\bibfnamefont {K.}~\bibnamefont {Kanoda}},\ }\href
  {\doibase 10.1103/PhysRevB.69.064511} {\bibfield  {journal} {\bibinfo
  {journal} {Phys. Rev. B}\ }\textbf {\bibinfo {volume} {69}},\ \bibinfo
  {pages} {64511} (\bibinfo {year} {2004})}\BibitemShut {NoStop}%
\bibitem [{\citenamefont {Furukawa}\ \emph {et~al.}(2015)\citenamefont
  {Furukawa}, \citenamefont {Miyagawa}, \citenamefont {Taniguchi},
  \citenamefont {Kato},\ and\ \citenamefont {Kanoda}}]{Furukawa2015}%
  \BibitemOpen
  \bibfield  {author} {\bibinfo {author} {\bibfnamefont {T.}~\bibnamefont
  {Furukawa}}, \bibinfo {author} {\bibfnamefont {K.}~\bibnamefont {Miyagawa}},
  \bibinfo {author} {\bibfnamefont {H.}~\bibnamefont {Taniguchi}}, \bibinfo
  {author} {\bibfnamefont {R.}~\bibnamefont {Kato}}, \ and\ \bibinfo {author}
  {\bibfnamefont {K.}~\bibnamefont {Kanoda}},\ }\href@noop {} {\bibfield
  {journal} {\bibinfo  {journal} {Nat. Phys.}\ }\textbf {\bibinfo {volume}
  {11}},\ \bibinfo {pages} {221} (\bibinfo {year} {2015})}\BibitemShut
  {NoStop}%
\bibitem [{\citenamefont {Kurosaki}\ \emph {et~al.}(2005)\citenamefont
  {Kurosaki}, \citenamefont {Shimizu}, \citenamefont {Miyagawa}, \citenamefont
  {Kanoda},\ and\ \citenamefont {Saito}}]{Kurosaki2005}%
  \BibitemOpen
  \bibfield  {author} {\bibinfo {author} {\bibfnamefont {Y.}~\bibnamefont
  {Kurosaki}}, \bibinfo {author} {\bibfnamefont {Y.}~\bibnamefont {Shimizu}},
  \bibinfo {author} {\bibfnamefont {K.}~\bibnamefont {Miyagawa}}, \bibinfo
  {author} {\bibfnamefont {K.}~\bibnamefont {Kanoda}}, \ and\ \bibinfo {author}
  {\bibfnamefont {G.}~\bibnamefont {Saito}},\ }\href
  {https://link.aps.org/doi/10.1103/PhysRevLett.95.177001} {\bibfield
  {journal} {\bibinfo  {journal} {Phys. Rev. Lett.}\ }\textbf {\bibinfo
  {volume} {95}},\ \bibinfo {pages} {177001} (\bibinfo {year}
  {2005})}\BibitemShut {NoStop}%
\bibitem [{\citenamefont {Shimizu}\ \emph {et~al.}(2003)\citenamefont
  {Shimizu}, \citenamefont {Miyagawa}, \citenamefont {Kanoda}, \citenamefont
  {Maesato},\ and\ \citenamefont {Saito}}]{Shimizu2003}%
  \BibitemOpen
  \bibfield  {author} {\bibinfo {author} {\bibfnamefont {Y.}~\bibnamefont
  {Shimizu}}, \bibinfo {author} {\bibfnamefont {K.}~\bibnamefont {Miyagawa}},
  \bibinfo {author} {\bibfnamefont {K.}~\bibnamefont {Kanoda}}, \bibinfo
  {author} {\bibfnamefont {M.}~\bibnamefont {Maesato}}, \ and\ \bibinfo
  {author} {\bibfnamefont {G.}~\bibnamefont {Saito}},\ }\href
  {https://link.aps.org/doi/10.1103/PhysRevLett.91.107001} {\bibfield
  {journal} {\bibinfo  {journal} {Phys. Rev. Lett.}\ }\textbf {\bibinfo
  {volume} {91}},\ \bibinfo {pages} {107001} (\bibinfo {year}
  {2003})}\BibitemShut {NoStop}%
\bibitem [{\citenamefont {Shimizu}\ \emph {et~al.}(2016)\citenamefont
  {Shimizu}, \citenamefont {Hiramatsu}, \citenamefont {Maesato}, \citenamefont
  {Otsuka}, \citenamefont {Yamochi}, \citenamefont {Ono}, \citenamefont {Itoh},
  \citenamefont {Yoshida}, \citenamefont {Takigawa}, \citenamefont {Yoshida},\
  and\ \citenamefont {Saito}}]{Shimizu2016}%
  \BibitemOpen
  \bibfield  {author} {\bibinfo {author} {\bibfnamefont {Y.}~\bibnamefont
  {Shimizu}}, \bibinfo {author} {\bibfnamefont {T.}~\bibnamefont {Hiramatsu}},
  \bibinfo {author} {\bibfnamefont {M.}~\bibnamefont {Maesato}}, \bibinfo
  {author} {\bibfnamefont {A.}~\bibnamefont {Otsuka}}, \bibinfo {author}
  {\bibfnamefont {H.}~\bibnamefont {Yamochi}}, \bibinfo {author} {\bibfnamefont
  {A.}~\bibnamefont {Ono}}, \bibinfo {author} {\bibfnamefont {M.}~\bibnamefont
  {Itoh}}, \bibinfo {author} {\bibfnamefont {M.}~\bibnamefont {Yoshida}},
  \bibinfo {author} {\bibfnamefont {M.}~\bibnamefont {Takigawa}}, \bibinfo
  {author} {\bibfnamefont {Y.}~\bibnamefont {Yoshida}}, \ and\ \bibinfo
  {author} {\bibfnamefont {G.}~\bibnamefont {Saito}},\ }\href
  {https://link.aps.org/doi/10.1103/PhysRevLett.117.107203} {\bibfield
  {journal} {\bibinfo  {journal} {Phys. Rev. Lett.}\ }\textbf {\bibinfo
  {volume} {117}},\ \bibinfo {pages} {107203} (\bibinfo {year}
  {2016})}\BibitemShut {NoStop}%
\bibitem [{\citenamefont {Itou}\ \emph {et~al.}(2017)\citenamefont {Itou},
  \citenamefont {Watanabe}, \citenamefont {Maegawa}, \citenamefont {Tajima},
  \citenamefont {Tajima}, \citenamefont {Kubo}, \citenamefont {Kato},\ and\
  \citenamefont {Kanoda}}]{Itou2017}%
  \BibitemOpen
  \bibfield  {author} {\bibinfo {author} {\bibfnamefont {T.}~\bibnamefont
  {Itou}}, \bibinfo {author} {\bibfnamefont {E.}~\bibnamefont {Watanabe}},
  \bibinfo {author} {\bibfnamefont {S.}~\bibnamefont {Maegawa}}, \bibinfo
  {author} {\bibfnamefont {A.}~\bibnamefont {Tajima}}, \bibinfo {author}
  {\bibfnamefont {N.}~\bibnamefont {Tajima}}, \bibinfo {author} {\bibfnamefont
  {K.}~\bibnamefont {Kubo}}, \bibinfo {author} {\bibfnamefont {R.}~\bibnamefont
  {Kato}}, \ and\ \bibinfo {author} {\bibfnamefont {K.}~\bibnamefont
  {Kanoda}},\ }\href
  {http://advances.sciencemag.org/content/3/8/e1601594.abstract} {\bibfield
  {journal} {\bibinfo  {journal} {Sci. Adv.}\ }\textbf {\bibinfo {volume}
  {3}},\ \bibinfo {pages} {e1601594} (\bibinfo {year} {2017})}\BibitemShut
  {NoStop}%
\bibitem [{\citenamefont {Li}\ \emph {et~al.}(2019)\citenamefont {Li},
  \citenamefont {Pustogow}, \citenamefont {Kato},\ and\ \citenamefont
  {Dressel}}]{Li2019}%
  \BibitemOpen
  \bibfield  {author} {\bibinfo {author} {\bibfnamefont {W.}~\bibnamefont
  {Li}}, \bibinfo {author} {\bibfnamefont {A.}~\bibnamefont {Pustogow}},
  \bibinfo {author} {\bibfnamefont {R.}~\bibnamefont {Kato}}, \ and\ \bibinfo
  {author} {\bibfnamefont {M.}~\bibnamefont {Dressel}},\ }\href {\doibase
  10.1103/PhysRevB.99.115137} {\bibfield  {journal} {\bibinfo  {journal} {Phys.
  Rev. B}\ }\textbf {\bibinfo {volume} {99}},\ \bibinfo {pages} {115137}
  (\bibinfo {year} {2019})}\BibitemShut {NoStop}%
\bibitem [{\citenamefont {Furukawa}\ \emph {et~al.}(2018)\citenamefont
  {Furukawa}, \citenamefont {Kobashi}, \citenamefont {Kurosaki}, \citenamefont
  {Miyagawa},\ and\ \citenamefont {Kanoda}}]{Furukawa2018}%
  \BibitemOpen
  \bibfield  {author} {\bibinfo {author} {\bibfnamefont {T.}~\bibnamefont
  {Furukawa}}, \bibinfo {author} {\bibfnamefont {K.}~\bibnamefont {Kobashi}},
  \bibinfo {author} {\bibfnamefont {Y.}~\bibnamefont {Kurosaki}}, \bibinfo
  {author} {\bibfnamefont {K.}~\bibnamefont {Miyagawa}}, \ and\ \bibinfo
  {author} {\bibfnamefont {K.}~\bibnamefont {Kanoda}},\ }\href {\doibase
  10.1038/s41467-017-02679-7} {\bibfield  {journal} {\bibinfo  {journal} {Nat.
  Commun.}\ }\textbf {\bibinfo {volume} {9}},\ \bibinfo {pages} {307} (\bibinfo
  {year} {2018})}\BibitemShut {NoStop}%
\bibitem [{\citenamefont {Pustogow}\ \emph
  {et~al.}(2018{\natexlab{a}})\citenamefont {Pustogow}, \citenamefont {Bories},
  \citenamefont {L{\"{o}}hle}, \citenamefont {R{\"{o}}sslhuber}, \citenamefont
  {Zhukova}, \citenamefont {Gorshunov}, \citenamefont {Tomi{\'{c}}},
  \citenamefont {Schlueter}, \citenamefont {H{\"{u}}bner}, \citenamefont
  {Hiramatsu}, \citenamefont {Yoshida}, \citenamefont {Saito}, \citenamefont
  {Kato}, \citenamefont {Lee}, \citenamefont {Dobrosavljevi{\'{c}}},
  \citenamefont {Fratini},\ and\ \citenamefont {Dressel}}]{Pustogow2018}%
  \BibitemOpen
  \bibfield  {author} {\bibinfo {author} {\bibfnamefont {A.}~\bibnamefont
  {Pustogow}}, \bibinfo {author} {\bibfnamefont {M.}~\bibnamefont {Bories}},
  \bibinfo {author} {\bibfnamefont {A.}~\bibnamefont {L{\"{o}}hle}}, \bibinfo
  {author} {\bibfnamefont {R.}~\bibnamefont {R{\"{o}}sslhuber}}, \bibinfo
  {author} {\bibfnamefont {E.}~\bibnamefont {Zhukova}}, \bibinfo {author}
  {\bibfnamefont {B.}~\bibnamefont {Gorshunov}}, \bibinfo {author}
  {\bibfnamefont {S.}~\bibnamefont {Tomi{\'{c}}}}, \bibinfo {author}
  {\bibfnamefont {J.~A.}\ \bibnamefont {Schlueter}}, \bibinfo {author}
  {\bibfnamefont {R.}~\bibnamefont {H{\"{u}}bner}}, \bibinfo {author}
  {\bibfnamefont {T.}~\bibnamefont {Hiramatsu}}, \bibinfo {author}
  {\bibfnamefont {Y.}~\bibnamefont {Yoshida}}, \bibinfo {author} {\bibfnamefont
  {G.}~\bibnamefont {Saito}}, \bibinfo {author} {\bibfnamefont
  {R.}~\bibnamefont {Kato}}, \bibinfo {author} {\bibfnamefont {T.-H.}\
  \bibnamefont {Lee}}, \bibinfo {author} {\bibfnamefont {V.}~\bibnamefont
  {Dobrosavljevi{\'{c}}}}, \bibinfo {author} {\bibfnamefont {S.}~\bibnamefont
  {Fratini}}, \ and\ \bibinfo {author} {\bibfnamefont {M.}~\bibnamefont
  {Dressel}},\ }\href {\doibase 10.1038/s41563-018-0140-3} {\bibfield
  {journal} {\bibinfo  {journal} {Nat. Mater.}\ }\textbf {\bibinfo {volume}
  {17}},\ \bibinfo {pages} {773} (\bibinfo {year}
  {2018}{\natexlab{a}})}\BibitemShut {NoStop}%
\bibitem [{\citenamefont {Terletska}\ \emph {et~al.}(2011)\citenamefont
  {Terletska}, \citenamefont {Vu{\v{c}}i{\v{c}}evi{\'{c}}}, \citenamefont
  {Tanaskovi{\'{c}}},\ and\ \citenamefont
  {Dobrosavljevi{\'{c}}}}]{Terletska2011}%
  \BibitemOpen
  \bibfield  {author} {\bibinfo {author} {\bibfnamefont {H.}~\bibnamefont
  {Terletska}}, \bibinfo {author} {\bibfnamefont {J.}~\bibnamefont
  {Vu{\v{c}}i{\v{c}}evi{\'{c}}}}, \bibinfo {author} {\bibfnamefont
  {D.}~\bibnamefont {Tanaskovi{\'{c}}}}, \ and\ \bibinfo {author}
  {\bibfnamefont {V.}~\bibnamefont {Dobrosavljevi{\'{c}}}},\ }\href
  {https://link.aps.org/doi/10.1103/PhysRevLett.107.026401} {\bibfield
  {journal} {\bibinfo  {journal} {Phys. Rev. Lett.}\ }\textbf {\bibinfo
  {volume} {107}},\ \bibinfo {pages} {26401} (\bibinfo {year}
  {2011})}\BibitemShut {NoStop}%
\bibitem [{\citenamefont {Vu{\v{c}}i{\v{c}}evi{\'{c}}}\ \emph
  {et~al.}(2013)\citenamefont {Vu{\v{c}}i{\v{c}}evi{\'{c}}}, \citenamefont
  {Terletska}, \citenamefont {Tanaskovi{\'{c}}},\ and\ \citenamefont
  {Dobrosavljevi{\'{c}}}}]{Vucicevic2013}%
  \BibitemOpen
  \bibfield  {author} {\bibinfo {author} {\bibfnamefont {J.}~\bibnamefont
  {Vu{\v{c}}i{\v{c}}evi{\'{c}}}}, \bibinfo {author} {\bibfnamefont
  {H.}~\bibnamefont {Terletska}}, \bibinfo {author} {\bibfnamefont
  {D.}~\bibnamefont {Tanaskovi{\'{c}}}}, \ and\ \bibinfo {author}
  {\bibfnamefont {V.}~\bibnamefont {Dobrosavljevi{\'{c}}}},\ }\href
  {https://link.aps.org/doi/10.1103/PhysRevB.88.075143} {\bibfield  {journal}
  {\bibinfo  {journal} {Phys. Rev. B}\ }\textbf {\bibinfo {volume} {88}},\
  \bibinfo {pages} {75143} (\bibinfo {year} {2013})}\BibitemShut {NoStop}%
\bibitem [{\citenamefont {Dobrosavljevi{\'{c}}}\ \emph
  {et~al.}(1997)\citenamefont {Dobrosavljevi{\'{c}}}, \citenamefont {Abrahams},
  \citenamefont {Miranda},\ and\ \citenamefont
  {Chakravarty}}]{Dobrosavljevic1997}%
  \BibitemOpen
  \bibfield  {author} {\bibinfo {author} {\bibfnamefont {V.}~\bibnamefont
  {Dobrosavljevi{\'{c}}}}, \bibinfo {author} {\bibfnamefont {E.}~\bibnamefont
  {Abrahams}}, \bibinfo {author} {\bibfnamefont {E.}~\bibnamefont {Miranda}}, \
  and\ \bibinfo {author} {\bibfnamefont {S.}~\bibnamefont {Chakravarty}},\
  }\href {\doibase 10.1103/PhysRevLett.79.455} {\bibfield  {journal} {\bibinfo
  {journal} {Phys. Rev. Lett.}\ }\textbf {\bibinfo {volume} {79}},\ \bibinfo
  {pages} {455} (\bibinfo {year} {1997})}\BibitemShut {NoStop}%
\bibitem [{\citenamefont {Radonji{\'{c}}}\ \emph {et~al.}(2012)\citenamefont
  {Radonji{\'{c}}}, \citenamefont {Tanaskovi{\'{c}}}, \citenamefont
  {Dobrosavljevi{\'{c}}}, \citenamefont {Haule},\ and\ \citenamefont
  {Kotliar}}]{Radonjic2012}%
  \BibitemOpen
  \bibfield  {author} {\bibinfo {author} {\bibfnamefont {M.~M.}\ \bibnamefont
  {Radonji{\'{c}}}}, \bibinfo {author} {\bibfnamefont {D.}~\bibnamefont
  {Tanaskovi{\'{c}}}}, \bibinfo {author} {\bibfnamefont {V.}~\bibnamefont
  {Dobrosavljevi{\'{c}}}}, \bibinfo {author} {\bibfnamefont {K.}~\bibnamefont
  {Haule}}, \ and\ \bibinfo {author} {\bibfnamefont {G.}~\bibnamefont
  {Kotliar}},\ }\href {https://link.aps.org/doi/10.1103/PhysRevB.85.085133}
  {\bibfield  {journal} {\bibinfo  {journal} {Phys. Rev. B}\ }\textbf {\bibinfo
  {volume} {85}},\ \bibinfo {pages} {85133} (\bibinfo {year}
  {2012})}\BibitemShut {NoStop}%
\bibitem [{\citenamefont {Deng}\ \emph {et~al.}(2013)\citenamefont {Deng},
  \citenamefont {Mravlje}, \citenamefont {{\v{Z}}itko}, \citenamefont
  {Ferrero}, \citenamefont {Kotliar},\ and\ \citenamefont
  {Georges}}]{Deng2013}%
  \BibitemOpen
  \bibfield  {author} {\bibinfo {author} {\bibfnamefont {X.}~\bibnamefont
  {Deng}}, \bibinfo {author} {\bibfnamefont {J.}~\bibnamefont {Mravlje}},
  \bibinfo {author} {\bibfnamefont {R.}~\bibnamefont {{\v{Z}}itko}}, \bibinfo
  {author} {\bibfnamefont {M.}~\bibnamefont {Ferrero}}, \bibinfo {author}
  {\bibfnamefont {G.}~\bibnamefont {Kotliar}}, \ and\ \bibinfo {author}
  {\bibfnamefont {A.}~\bibnamefont {Georges}},\ }\href
  {https://link.aps.org/doi/10.1103/PhysRevLett.110.086401} {\bibfield
  {journal} {\bibinfo  {journal} {Phys. Rev. Lett.}\ }\textbf {\bibinfo
  {volume} {110}},\ \bibinfo {pages} {86401} (\bibinfo {year}
  {2013})}\BibitemShut {NoStop}%
\bibitem [{\citenamefont {Lefebvre}\ \emph {et~al.}(2000)\citenamefont
  {Lefebvre}, \citenamefont {Wzietek}, \citenamefont {Brown}, \citenamefont
  {Bourbonnais}, \citenamefont {J{\'{e}}rome}, \citenamefont
  {M{\'{e}}zi{\`{e}}re}, \citenamefont {Fourmigu{\'{e}}},\ and\ \citenamefont
  {Batail}}]{Lefebvre2000}%
  \BibitemOpen
  \bibfield  {author} {\bibinfo {author} {\bibfnamefont {S.}~\bibnamefont
  {Lefebvre}}, \bibinfo {author} {\bibfnamefont {P.}~\bibnamefont {Wzietek}},
  \bibinfo {author} {\bibfnamefont {S.}~\bibnamefont {Brown}}, \bibinfo
  {author} {\bibfnamefont {C.}~\bibnamefont {Bourbonnais}}, \bibinfo {author}
  {\bibfnamefont {D.}~\bibnamefont {J{\'{e}}rome}}, \bibinfo {author}
  {\bibfnamefont {C.}~\bibnamefont {M{\'{e}}zi{\`{e}}re}}, \bibinfo {author}
  {\bibfnamefont {M.}~\bibnamefont {Fourmigu{\'{e}}}}, \ and\ \bibinfo {author}
  {\bibfnamefont {P.}~\bibnamefont {Batail}},\ }\href@noop {} {\bibfield
  {journal} {\bibinfo  {journal} {Phys. Rev. Lett.}\ }\textbf {\bibinfo
  {volume} {85}},\ \bibinfo {pages} {5420} (\bibinfo {year}
  {2000})}\BibitemShut {NoStop}%
\bibitem [{\citenamefont {Sasaki}\ \emph {et~al.}(2004)\citenamefont {Sasaki},
  \citenamefont {Yoneyama}, \citenamefont {Kobayashi}, \citenamefont
  {Ikemoto},\ and\ \citenamefont {Kimura}}]{Sasaki2004}%
  \BibitemOpen
  \bibfield  {author} {\bibinfo {author} {\bibfnamefont {T.}~\bibnamefont
  {Sasaki}}, \bibinfo {author} {\bibfnamefont {N.}~\bibnamefont {Yoneyama}},
  \bibinfo {author} {\bibfnamefont {N.}~\bibnamefont {Kobayashi}}, \bibinfo
  {author} {\bibfnamefont {Y.}~\bibnamefont {Ikemoto}}, \ and\ \bibinfo
  {author} {\bibfnamefont {H.}~\bibnamefont {Kimura}},\ }\href {\doibase
  10.1103/PhysRevLett.92.227001} {\bibfield  {journal} {\bibinfo  {journal}
  {Phys. Rev. Lett.}\ }\textbf {\bibinfo {volume} {92}},\ \bibinfo {pages}
  {227001} (\bibinfo {year} {2004})}\BibitemShut {NoStop}%
\bibitem [{\citenamefont {Qazilbash}\ \emph {et~al.}(2007)\citenamefont
  {Qazilbash}, \citenamefont {Brehm}, \citenamefont {Chae}, \citenamefont {Ho},
  \citenamefont {Andreev}, \citenamefont {Kim}, \citenamefont {Yun},
  \citenamefont {Balatsky}, \citenamefont {Maple}, \citenamefont {Keilmann},
  \citenamefont {Kim},\ and\ \citenamefont {Basov}}]{Qazilbash2007}%
  \BibitemOpen
  \bibfield  {author} {\bibinfo {author} {\bibfnamefont {M.~M.}\ \bibnamefont
  {Qazilbash}}, \bibinfo {author} {\bibfnamefont {M.}~\bibnamefont {Brehm}},
  \bibinfo {author} {\bibfnamefont {B.-G.}\ \bibnamefont {Chae}}, \bibinfo
  {author} {\bibfnamefont {P.-C.}\ \bibnamefont {Ho}}, \bibinfo {author}
  {\bibfnamefont {G.~O.}\ \bibnamefont {Andreev}}, \bibinfo {author}
  {\bibfnamefont {B.-J.}\ \bibnamefont {Kim}}, \bibinfo {author} {\bibfnamefont
  {S.~J.}\ \bibnamefont {Yun}}, \bibinfo {author} {\bibfnamefont {A.~V.}\
  \bibnamefont {Balatsky}}, \bibinfo {author} {\bibfnamefont {M.~B.}\
  \bibnamefont {Maple}}, \bibinfo {author} {\bibfnamefont {F.}~\bibnamefont
  {Keilmann}}, \bibinfo {author} {\bibfnamefont {H.-T.}\ \bibnamefont {Kim}}, \
  and\ \bibinfo {author} {\bibfnamefont {D.~N.}\ \bibnamefont {Basov}},\ }\href
  {\doibase 10.1126/science.1150124} {\bibfield  {journal} {\bibinfo  {journal}
  {Science}\ }\textbf {\bibinfo {volume} {318}},\ \bibinfo {pages} {1750}
  (\bibinfo {year} {2007})}\BibitemShut {NoStop}%
\bibitem [{\citenamefont {Huffman}\ \emph {et~al.}(2018)\citenamefont
  {Huffman}, \citenamefont {Lahneman}, \citenamefont {Wang}, \citenamefont
  {Slusar}, \citenamefont {Kim}, \citenamefont {Kim},\ and\ \citenamefont
  {Qazilbash}}]{Huffman2018}%
  \BibitemOpen
  \bibfield  {author} {\bibinfo {author} {\bibfnamefont {T.~J.}\ \bibnamefont
  {Huffman}}, \bibinfo {author} {\bibfnamefont {D.~J.}\ \bibnamefont
  {Lahneman}}, \bibinfo {author} {\bibfnamefont {S.~L.}\ \bibnamefont {Wang}},
  \bibinfo {author} {\bibfnamefont {T.}~\bibnamefont {Slusar}}, \bibinfo
  {author} {\bibfnamefont {B.-J.}\ \bibnamefont {Kim}}, \bibinfo {author}
  {\bibfnamefont {H.-T.}\ \bibnamefont {Kim}}, \ and\ \bibinfo {author}
  {\bibfnamefont {M.~M.}\ \bibnamefont {Qazilbash}},\ }\href {\doibase
  10.1103/PhysRevB.97.085146} {\bibfield  {journal} {\bibinfo  {journal} {Phys.
  Rev. B}\ }\textbf {\bibinfo {volume} {97}},\ \bibinfo {pages} {085146}
  (\bibinfo {year} {2018})}\BibitemShut {NoStop}%
\bibitem [{\citenamefont {McLeod}\ \emph {et~al.}(2016)\citenamefont {McLeod},
  \citenamefont {van Heumen}, \citenamefont {Ramirez}, \citenamefont {Wang},
  \citenamefont {Saerbeck}, \citenamefont {Guenon}, \citenamefont {Goldflam},
  \citenamefont {Anderegg}, \citenamefont {Kelly}, \citenamefont {Mueller},
  \citenamefont {Liu}, \citenamefont {Schuller},\ and\ \citenamefont
  {Basov}}]{McLeod2016}%
  \BibitemOpen
  \bibfield  {author} {\bibinfo {author} {\bibfnamefont {A.~S.}\ \bibnamefont
  {McLeod}}, \bibinfo {author} {\bibfnamefont {E.}~\bibnamefont {van Heumen}},
  \bibinfo {author} {\bibfnamefont {J.~G.}\ \bibnamefont {Ramirez}}, \bibinfo
  {author} {\bibfnamefont {S.}~\bibnamefont {Wang}}, \bibinfo {author}
  {\bibfnamefont {T.}~\bibnamefont {Saerbeck}}, \bibinfo {author}
  {\bibfnamefont {S.}~\bibnamefont {Guenon}}, \bibinfo {author} {\bibfnamefont
  {M.}~\bibnamefont {Goldflam}}, \bibinfo {author} {\bibfnamefont
  {L.}~\bibnamefont {Anderegg}}, \bibinfo {author} {\bibfnamefont
  {P.}~\bibnamefont {Kelly}}, \bibinfo {author} {\bibfnamefont
  {A.}~\bibnamefont {Mueller}}, \bibinfo {author} {\bibfnamefont {M.~K.}\
  \bibnamefont {Liu}}, \bibinfo {author} {\bibfnamefont {I.~K.}\ \bibnamefont
  {Schuller}}, \ and\ \bibinfo {author} {\bibfnamefont {D.~N.}\ \bibnamefont
  {Basov}},\ }\href {\doibase 10.1038/nphys3882} {\bibfield  {journal}
  {\bibinfo  {journal} {Nat. Phys.}\ }\textbf {\bibinfo {volume} {13}},\
  \bibinfo {pages} {80} (\bibinfo {year} {2016})}\BibitemShut {NoStop}%
\bibitem [{\citenamefont {Lupi}\ \emph {et~al.}(2010)\citenamefont {Lupi},
  \citenamefont {Baldassarre}, \citenamefont {Mansart}, \citenamefont
  {Perucchi}, \citenamefont {Barinov}, \citenamefont {Dudin}, \citenamefont
  {Papalazarou}, \citenamefont {Rodolakis}, \citenamefont {Rueff},
  \citenamefont {Iti{\'{e}}}, \citenamefont {Ravy}, \citenamefont {Nicoletti},
  \citenamefont {Postorino}, \citenamefont {Hansmann}, \citenamefont {Parragh},
  \citenamefont {Toschi}, \citenamefont {Saha-Dasgupta}, \citenamefont
  {Andersen}, \citenamefont {Sangiovanni}, \citenamefont {Held},\ and\
  \citenamefont {Marsi}}]{Lupi2010}%
  \BibitemOpen
  \bibfield  {author} {\bibinfo {author} {\bibfnamefont {S.}~\bibnamefont
  {Lupi}}, \bibinfo {author} {\bibfnamefont {L.}~\bibnamefont {Baldassarre}},
  \bibinfo {author} {\bibfnamefont {B.}~\bibnamefont {Mansart}}, \bibinfo
  {author} {\bibfnamefont {A.}~\bibnamefont {Perucchi}}, \bibinfo {author}
  {\bibfnamefont {A.}~\bibnamefont {Barinov}}, \bibinfo {author} {\bibfnamefont
  {P.}~\bibnamefont {Dudin}}, \bibinfo {author} {\bibfnamefont
  {E.}~\bibnamefont {Papalazarou}}, \bibinfo {author} {\bibfnamefont
  {F.}~\bibnamefont {Rodolakis}}, \bibinfo {author} {\bibfnamefont {J.~P.}\
  \bibnamefont {Rueff}}, \bibinfo {author} {\bibfnamefont {J.~P.}\ \bibnamefont
  {Iti{\'{e}}}}, \bibinfo {author} {\bibfnamefont {S.}~\bibnamefont {Ravy}},
  \bibinfo {author} {\bibfnamefont {D.}~\bibnamefont {Nicoletti}}, \bibinfo
  {author} {\bibfnamefont {P.}~\bibnamefont {Postorino}}, \bibinfo {author}
  {\bibfnamefont {P.}~\bibnamefont {Hansmann}}, \bibinfo {author}
  {\bibfnamefont {N.}~\bibnamefont {Parragh}}, \bibinfo {author} {\bibfnamefont
  {A.}~\bibnamefont {Toschi}}, \bibinfo {author} {\bibfnamefont
  {T.}~\bibnamefont {Saha-Dasgupta}}, \bibinfo {author} {\bibfnamefont {O.~K.}\
  \bibnamefont {Andersen}}, \bibinfo {author} {\bibfnamefont {G.}~\bibnamefont
  {Sangiovanni}}, \bibinfo {author} {\bibfnamefont {K.}~\bibnamefont {Held}}, \
  and\ \bibinfo {author} {\bibfnamefont {M.}~\bibnamefont {Marsi}},\ }\href
  {\doibase 10.1038/ncomms1109} {\bibfield  {journal} {\bibinfo  {journal}
  {Nat. Commun.}\ }\textbf {\bibinfo {volume} {1}},\ \bibinfo {pages} {105}
  (\bibinfo {year} {2010})}\BibitemShut {NoStop}%
\bibitem [{\citenamefont {Post}\ \emph {et~al.}(2018)\citenamefont {Post},
  \citenamefont {McLeod}, \citenamefont {Hepting}, \citenamefont {Bluschke},
  \citenamefont {Wang}, \citenamefont {Cristiani}, \citenamefont {Logvenov},
  \citenamefont {Charnukha}, \citenamefont {Ni}, \citenamefont {Radhakrishnan},
  \citenamefont {Minola}, \citenamefont {Pasupathy}, \citenamefont {Boris},
  \citenamefont {Benckiser}, \citenamefont {Dahmen}, \citenamefont {Carlson},
  \citenamefont {Keimer},\ and\ \citenamefont {Basov}}]{Post2018}%
  \BibitemOpen
  \bibfield  {author} {\bibinfo {author} {\bibfnamefont {K.~W.}\ \bibnamefont
  {Post}}, \bibinfo {author} {\bibfnamefont {A.~S.}\ \bibnamefont {McLeod}},
  \bibinfo {author} {\bibfnamefont {M.}~\bibnamefont {Hepting}}, \bibinfo
  {author} {\bibfnamefont {M.}~\bibnamefont {Bluschke}}, \bibinfo {author}
  {\bibfnamefont {Y.}~\bibnamefont {Wang}}, \bibinfo {author} {\bibfnamefont
  {G.}~\bibnamefont {Cristiani}}, \bibinfo {author} {\bibfnamefont
  {G.}~\bibnamefont {Logvenov}}, \bibinfo {author} {\bibfnamefont
  {A.}~\bibnamefont {Charnukha}}, \bibinfo {author} {\bibfnamefont {G.~X.}\
  \bibnamefont {Ni}}, \bibinfo {author} {\bibfnamefont {P.}~\bibnamefont
  {Radhakrishnan}}, \bibinfo {author} {\bibfnamefont {M.}~\bibnamefont
  {Minola}}, \bibinfo {author} {\bibfnamefont {A.}~\bibnamefont {Pasupathy}},
  \bibinfo {author} {\bibfnamefont {A.~V.}\ \bibnamefont {Boris}}, \bibinfo
  {author} {\bibfnamefont {E.}~\bibnamefont {Benckiser}}, \bibinfo {author}
  {\bibfnamefont {K.~A.}\ \bibnamefont {Dahmen}}, \bibinfo {author}
  {\bibfnamefont {E.~W.}\ \bibnamefont {Carlson}}, \bibinfo {author}
  {\bibfnamefont {B.}~\bibnamefont {Keimer}}, \ and\ \bibinfo {author}
  {\bibfnamefont {D.~N.}\ \bibnamefont {Basov}},\ }\href {\doibase
  10.1038/s41567-018-0201-1} {\bibfield  {journal} {\bibinfo  {journal} {Nat.
  Phys.}\ }\textbf {\bibinfo {volume} {14}},\ \bibinfo {pages} {1056} (\bibinfo
  {year} {2018})}\BibitemShut {NoStop}%
\bibitem [{\citenamefont {Pustogow}\ \emph
  {et~al.}(2018{\natexlab{b}})\citenamefont {Pustogow}, \citenamefont {McLeod},
  \citenamefont {Saito}, \citenamefont {Basov},\ and\ \citenamefont
  {Dressel}}]{Pustogow2018s}%
  \BibitemOpen
  \bibfield  {author} {\bibinfo {author} {\bibfnamefont {A.}~\bibnamefont
  {Pustogow}}, \bibinfo {author} {\bibfnamefont {A.~S.}\ \bibnamefont
  {McLeod}}, \bibinfo {author} {\bibfnamefont {Y.}~\bibnamefont {Saito}},
  \bibinfo {author} {\bibfnamefont {D.~N.}\ \bibnamefont {Basov}}, \ and\
  \bibinfo {author} {\bibfnamefont {M.}~\bibnamefont {Dressel}},\ }\href
  {\doibase 10.1126/sciadv.aau9123} {\bibfield  {journal} {\bibinfo  {journal}
  {Sci. Adv.}\ }\textbf {\bibinfo {volume} {4}},\ \bibinfo {pages} {eaau9123}
  (\bibinfo {year} {2018}{\natexlab{b}})}\BibitemShut {NoStop}%
\bibitem [{\citenamefont {Abdel-Jawad}\ \emph {et~al.}(2010)\citenamefont
  {Abdel-Jawad}, \citenamefont {Terasaki}, \citenamefont {Sasaki},
  \citenamefont {Yoneyama}, \citenamefont {Kobayashi}, \citenamefont {Uesu},\
  and\ \citenamefont {Hotta}}]{Abdel-Jawad2010}%
  \BibitemOpen
  \bibfield  {author} {\bibinfo {author} {\bibfnamefont {M.}~\bibnamefont
  {Abdel-Jawad}}, \bibinfo {author} {\bibfnamefont {I.}~\bibnamefont
  {Terasaki}}, \bibinfo {author} {\bibfnamefont {T.}~\bibnamefont {Sasaki}},
  \bibinfo {author} {\bibfnamefont {N.}~\bibnamefont {Yoneyama}}, \bibinfo
  {author} {\bibfnamefont {N.}~\bibnamefont {Kobayashi}}, \bibinfo {author}
  {\bibfnamefont {Y.}~\bibnamefont {Uesu}}, \ and\ \bibinfo {author}
  {\bibfnamefont {C.}~\bibnamefont {Hotta}},\ }\href {\doibase
  10.1103/PhysRevB.82.125119} {\bibfield  {journal} {\bibinfo  {journal} {Phys.
  Rev. B}\ }\textbf {\bibinfo {volume} {82}},\ \bibinfo {pages} {125119}
  (\bibinfo {year} {2010})}\BibitemShut {NoStop}%
\bibitem [{\citenamefont {Pinteri{\'{c}}}\ \emph {et~al.}(2014)\citenamefont
  {Pinteri{\'{c}}}, \citenamefont {{\v{C}}ulo}, \citenamefont {Milat},
  \citenamefont {Basleti{\'{c}}}, \citenamefont {Korin-Hamzi{\'{c}}},
  \citenamefont {Tafra}, \citenamefont {Hamzi{\'{c}}}, \citenamefont {Ivek},
  \citenamefont {Peterseim}, \citenamefont {Miyagawa}, \citenamefont {Kanoda},
  \citenamefont {Schlueter}, \citenamefont {Dressel},\ and\ \citenamefont
  {Tomi{\'{c}}}}]{Pinteric2014}%
  \BibitemOpen
  \bibfield  {author} {\bibinfo {author} {\bibfnamefont {M.}~\bibnamefont
  {Pinteri{\'{c}}}}, \bibinfo {author} {\bibfnamefont {M.}~\bibnamefont
  {{\v{C}}ulo}}, \bibinfo {author} {\bibfnamefont {O.}~\bibnamefont {Milat}},
  \bibinfo {author} {\bibfnamefont {M.}~\bibnamefont {Basleti{\'{c}}}},
  \bibinfo {author} {\bibfnamefont {B.}~\bibnamefont {Korin-Hamzi{\'{c}}}},
  \bibinfo {author} {\bibfnamefont {E.}~\bibnamefont {Tafra}}, \bibinfo
  {author} {\bibfnamefont {A.}~\bibnamefont {Hamzi{\'{c}}}}, \bibinfo {author}
  {\bibfnamefont {T.}~\bibnamefont {Ivek}}, \bibinfo {author} {\bibfnamefont
  {T.}~\bibnamefont {Peterseim}}, \bibinfo {author} {\bibfnamefont
  {K.}~\bibnamefont {Miyagawa}}, \bibinfo {author} {\bibfnamefont
  {K.}~\bibnamefont {Kanoda}}, \bibinfo {author} {\bibfnamefont {J.~A.}\
  \bibnamefont {Schlueter}}, \bibinfo {author} {\bibfnamefont {M.}~\bibnamefont
  {Dressel}}, \ and\ \bibinfo {author} {\bibfnamefont {S.}~\bibnamefont
  {Tomi{\'{c}}}},\ }\href {https://link.aps.org/doi/10.1103/PhysRevB.90.195139}
  {\bibfield  {journal} {\bibinfo  {journal} {Phys. Rev. B}\ }\textbf {\bibinfo
  {volume} {90}},\ \bibinfo {pages} {195139} (\bibinfo {year}
  {2014})}\BibitemShut {NoStop}%
\bibitem [{\citenamefont {K{\'{e}}zsm{\'{a}}rki}\ \emph
  {et~al.}(2006)\citenamefont {K{\'{e}}zsm{\'{a}}rki}, \citenamefont {Shimizu},
  \citenamefont {Mih{\'{a}}ly}, \citenamefont {Tokura}, \citenamefont
  {Kanoda},\ and\ \citenamefont {Saito}}]{Kezsmarki2006}%
  \BibitemOpen
  \bibfield  {author} {\bibinfo {author} {\bibfnamefont {I.}~\bibnamefont
  {K{\'{e}}zsm{\'{a}}rki}}, \bibinfo {author} {\bibfnamefont {Y.}~\bibnamefont
  {Shimizu}}, \bibinfo {author} {\bibfnamefont {G.}~\bibnamefont
  {Mih{\'{a}}ly}}, \bibinfo {author} {\bibfnamefont {Y.}~\bibnamefont
  {Tokura}}, \bibinfo {author} {\bibfnamefont {K.}~\bibnamefont {Kanoda}}, \
  and\ \bibinfo {author} {\bibfnamefont {G.}~\bibnamefont {Saito}},\ }\href
  {https://link.aps.org/doi/10.1103/PhysRevB.74.201101} {\bibfield  {journal}
  {\bibinfo  {journal} {Phys. Rev. B}\ }\textbf {\bibinfo {volume} {74}},\
  \bibinfo {pages} {201101} (\bibinfo {year} {2006})}\BibitemShut {NoStop}%
\bibitem [{\citenamefont {Kanoda}\ and\ \citenamefont
  {Kato}(2011)}]{Kanoda2011}%
  \BibitemOpen
  \bibfield  {author} {\bibinfo {author} {\bibfnamefont {K.}~\bibnamefont
  {Kanoda}}\ and\ \bibinfo {author} {\bibfnamefont {R.}~\bibnamefont {Kato}},\
  }\href {\doibase 10.1146/annurev-conmatphys-062910-140521} {\bibfield
  {journal} {\bibinfo  {journal} {Annu. Rev. Condens. Matter Phys.}\ }\textbf
  {\bibinfo {volume} {2}},\ \bibinfo {pages} {167} (\bibinfo {year}
  {2011})}\BibitemShut {NoStop}%
\bibitem [{\citenamefont {Zhou}\ \emph {et~al.}(2017)\citenamefont {Zhou},
  \citenamefont {Kanoda},\ and\ \citenamefont {Ng}}]{Zhou2017}%
  \BibitemOpen
  \bibfield  {author} {\bibinfo {author} {\bibfnamefont {Y.}~\bibnamefont
  {Zhou}}, \bibinfo {author} {\bibfnamefont {K.}~\bibnamefont {Kanoda}}, \ and\
  \bibinfo {author} {\bibfnamefont {T.-K.}\ \bibnamefont {Ng}},\ }\href
  {https://link.aps.org/doi/10.1103/RevModPhys.89.025003} {\bibfield  {journal}
  {\bibinfo  {journal} {Rev. Mod. Phys.}\ }\textbf {\bibinfo {volume} {89}},\
  \bibinfo {pages} {25003} (\bibinfo {year} {2017})}\BibitemShut {NoStop}%
\bibitem [{\citenamefont {{\v{C}}ulo}\ \emph {et~al.}(2019)\citenamefont
  {{\v{C}}ulo}, \citenamefont {Tafra}, \citenamefont {Mihaljevi{\'{c}}},
  \citenamefont {Basleti{\'{c}}}, \citenamefont {Kuve{\v{z}}di{\'{c}}},
  \citenamefont {Ivek}, \citenamefont {Hamzi{\'{c}}}, \citenamefont
  {Tomi{\'{c}}}, \citenamefont {Hiramatsu}, \citenamefont {Yoshida},
  \citenamefont {Saito}, \citenamefont {Schlueter}, \citenamefont {Dressel},\
  and\ \citenamefont {Korin-Hamzi{\'{c}}}}]{Culo2019}%
  \BibitemOpen
  \bibfield  {author} {\bibinfo {author} {\bibfnamefont {M.}~\bibnamefont
  {{\v{C}}ulo}}, \bibinfo {author} {\bibfnamefont {E.}~\bibnamefont {Tafra}},
  \bibinfo {author} {\bibfnamefont {B.}~\bibnamefont {Mihaljevi{\'{c}}}},
  \bibinfo {author} {\bibfnamefont {M.}~\bibnamefont {Basleti{\'{c}}}},
  \bibinfo {author} {\bibfnamefont {M.}~\bibnamefont {Kuve{\v{z}}di{\'{c}}}},
  \bibinfo {author} {\bibfnamefont {T.}~\bibnamefont {Ivek}}, \bibinfo {author}
  {\bibfnamefont {A.}~\bibnamefont {Hamzi{\'{c}}}}, \bibinfo {author}
  {\bibfnamefont {S.}~\bibnamefont {Tomi{\'{c}}}}, \bibinfo {author}
  {\bibfnamefont {T.}~\bibnamefont {Hiramatsu}}, \bibinfo {author}
  {\bibfnamefont {Y.}~\bibnamefont {Yoshida}}, \bibinfo {author} {\bibfnamefont
  {G.}~\bibnamefont {Saito}}, \bibinfo {author} {\bibfnamefont {J.~A.}\
  \bibnamefont {Schlueter}}, \bibinfo {author} {\bibfnamefont {M.}~\bibnamefont
  {Dressel}}, \ and\ \bibinfo {author} {\bibfnamefont {B.}~\bibnamefont
  {Korin-Hamzi{\'{c}}}},\ }\href {\doibase 10.1103/PhysRevB.99.045114}
  {\bibfield  {journal} {\bibinfo  {journal} {Phys. Rev. B}\ }\textbf {\bibinfo
  {volume} {99}},\ \bibinfo {pages} {45114} (\bibinfo {year}
  {2019})}\BibitemShut {NoStop}%
\bibitem [{Note1()}]{Note1}%
  \BibitemOpen
  \bibinfo {note} {The superconducting state at $T\approx 4$~K \cite
  {Kurosaki2005,Furukawa2018} is below the temperature accessible to us
  here.}\BibitemShut {Stop}%
\bibitem [{Note2()}]{Note2}%
  \BibitemOpen
  \bibinfo {note} {BEDT-TTF stands for bis-ethylene-dithio-tetra\protect
  \-thia\protect \-fulvalene. Substituting two of the inner sulfur atoms by
  selenium leads to bis-ethylene\protect \-dithio-di\protect
  \-selenium-di\protect \-thia\protect \-fulvalene, abbreviated BEDT-STF \cite
  {Saito2019}.}\BibitemShut {Stop}%
\bibitem [{\citenamefont {R{\"{o}}sslhuber}\ \emph {et~al.}()\citenamefont
  {R{\"{o}}sslhuber}, \citenamefont {Pustogow}, \citenamefont {Uykur},
  \citenamefont {B{\"{o}}hme}, \citenamefont {L{\"{o}}hle}, \citenamefont
  {H{\"{u}}bner}, \citenamefont {Schlueter}, \citenamefont {Tan}, \citenamefont
  {Dobrosavljevi\'c},\ and\ \citenamefont {Dressel}}]{Rosslhuber2019}%
  \BibitemOpen
  \bibfield  {author} {\bibinfo {author} {\bibfnamefont {R.}~\bibnamefont
  {R{\"{o}}sslhuber}}, \bibinfo {author} {\bibfnamefont {A.}~\bibnamefont
  {Pustogow}}, \bibinfo {author} {\bibfnamefont {E.}~\bibnamefont {Uykur}},
  \bibinfo {author} {\bibfnamefont {A.}~\bibnamefont {B{\"{o}}hme}}, \bibinfo
  {author} {\bibfnamefont {A.}~\bibnamefont {L{\"{o}}hle}}, \bibinfo {author}
  {\bibfnamefont {R.}~\bibnamefont {H{\"{u}}bner}}, \bibinfo {author}
  {\bibfnamefont {J.~A.}\ \bibnamefont {Schlueter}}, \bibinfo {author}
  {\bibfnamefont {Y.}~\bibnamefont {Tan}}, \bibinfo {author} {\bibfnamefont
  {V.}~\bibnamefont {Dobrosavljevi\'c}}, \ and\ \bibinfo {author}
  {\bibfnamefont {M.}~\bibnamefont {Dressel}},\ }\href@noop {} {}\bibinfo
  {note} {ArXiv:1911.12273}\BibitemShut {NoStop}%
\bibitem [{\citenamefont {Saito}\ \emph {et~al.}(2018)\citenamefont {Saito},
  \citenamefont {Minamidate}, \citenamefont {Kawamoto}, \citenamefont
  {Matsunaga},\ and\ \citenamefont {Nomura}}]{Saito2018}%
  \BibitemOpen
  \bibfield  {author} {\bibinfo {author} {\bibfnamefont {Y.}~\bibnamefont
  {Saito}}, \bibinfo {author} {\bibfnamefont {T.}~\bibnamefont {Minamidate}},
  \bibinfo {author} {\bibfnamefont {A.}~\bibnamefont {Kawamoto}}, \bibinfo
  {author} {\bibfnamefont {N.}~\bibnamefont {Matsunaga}}, \ and\ \bibinfo
  {author} {\bibfnamefont {K.}~\bibnamefont {Nomura}},\ }\href {\doibase
  10.1103/PhysRevB.98.205141} {\bibfield  {journal} {\bibinfo  {journal} {Phys.
  Rev. B}\ }\textbf {\bibinfo {volume} {98}},\ \bibinfo {pages} {205141}
  (\bibinfo {year} {2018})}\BibitemShut {NoStop}%
\bibitem [{\citenamefont {Saito}\ \emph {et~al.}()\citenamefont {Saito},
  \citenamefont {R{\"o}sslhuber}, \citenamefont {L{\"o}hle}, \citenamefont
  {Sanz~Alonso}, \citenamefont {Wenzel}, \citenamefont {Kawamoto},
  \citenamefont {Pustogow},\ and\ \citenamefont {Dressel}}]{Saito2019}%
  \BibitemOpen
  \bibfield  {author} {\bibinfo {author} {\bibfnamefont {Y.}~\bibnamefont
  {Saito}}, \bibinfo {author} {\bibfnamefont {R.}~\bibnamefont
  {R{\"o}sslhuber}}, \bibinfo {author} {\bibfnamefont {A.}~\bibnamefont
  {L{\"o}hle}}, \bibinfo {author} {\bibfnamefont {M.}~\bibnamefont
  {Sanz~Alonso}}, \bibinfo {author} {\bibfnamefont {M.}~\bibnamefont {Wenzel}},
  \bibinfo {author} {\bibfnamefont {A.}~\bibnamefont {Kawamoto}}, \bibinfo
  {author} {\bibfnamefont {A.}~\bibnamefont {Pustogow}}, \ and\ \bibinfo
  {author} {\bibfnamefont {M.}~\bibnamefont {Dressel}},\ }\href@noop {} {\
  }\bibinfo {note} {ArXiv:1911.06766}\BibitemShut {NoStop}%
\bibitem [{Note3()}]{Note3}%
  \BibitemOpen
  \bibinfo {note} {Comparison of the results in Fig.~\ref {fig:sigma-eps_T}(j)
  with optical data measured on the same substitution yields fair agreement of
  the metallic values $\epsilon _1 < 0$~\cite {Saito2019}. Technical details of
  the dielectric experiments can be found in Ref.~\cite
  {Rosslhuber2019}.}\BibitemShut {Stop}%
\bibitem [{\citenamefont {Faltermeier}\ \emph {et~al.}(2007)\citenamefont
  {Faltermeier}, \citenamefont {Barz}, \citenamefont {Dumm}, \citenamefont
  {Dressel}, \citenamefont {Drichko}, \citenamefont {Petrov}, \citenamefont
  {Semkin}, \citenamefont {Vlasova}, \citenamefont {Me{\'{z}}i{\`{e}}re},\ and\
  \citenamefont {Batail}}]{Faltermeier2007}%
  \BibitemOpen
  \bibfield  {author} {\bibinfo {author} {\bibfnamefont {D.}~\bibnamefont
  {Faltermeier}}, \bibinfo {author} {\bibfnamefont {J.}~\bibnamefont {Barz}},
  \bibinfo {author} {\bibfnamefont {M.}~\bibnamefont {Dumm}}, \bibinfo {author}
  {\bibfnamefont {M.}~\bibnamefont {Dressel}}, \bibinfo {author} {\bibfnamefont
  {N.}~\bibnamefont {Drichko}}, \bibinfo {author} {\bibfnamefont
  {B.}~\bibnamefont {Petrov}}, \bibinfo {author} {\bibfnamefont
  {V.}~\bibnamefont {Semkin}}, \bibinfo {author} {\bibfnamefont
  {R.}~\bibnamefont {Vlasova}}, \bibinfo {author} {\bibfnamefont
  {C.}~\bibnamefont {Me{\'{z}}i{\`{e}}re}}, \ and\ \bibinfo {author}
  {\bibfnamefont {P.}~\bibnamefont {Batail}},\ }\href
  {https://link.aps.org/doi/10.1103/PhysRevB.76.165113} {\bibfield  {journal}
  {\bibinfo  {journal} {Phys. Rev. B}\ }\textbf {\bibinfo {volume} {76}},\
  \bibinfo {pages} {165113} (\bibinfo {year} {2007})}\BibitemShut {NoStop}%
\bibitem [{\citenamefont {Merino}\ \emph {et~al.}(2008)\citenamefont {Merino},
  \citenamefont {Dumm}, \citenamefont {Drichko}, \citenamefont {Dressel},\ and\
  \citenamefont {McKenzie}}]{Merino2008}%
  \BibitemOpen
  \bibfield  {author} {\bibinfo {author} {\bibfnamefont {J.}~\bibnamefont
  {Merino}}, \bibinfo {author} {\bibfnamefont {M.}~\bibnamefont {Dumm}},
  \bibinfo {author} {\bibfnamefont {N.}~\bibnamefont {Drichko}}, \bibinfo
  {author} {\bibfnamefont {M.}~\bibnamefont {Dressel}}, \ and\ \bibinfo
  {author} {\bibfnamefont {R.~H.}\ \bibnamefont {McKenzie}},\ }\href
  {https://link.aps.org/doi/10.1103/PhysRevLett.100.086404} {\bibfield
  {journal} {\bibinfo  {journal} {Phys. Rev. Lett.}\ }\textbf {\bibinfo
  {volume} {100}},\ \bibinfo {pages} {86404} (\bibinfo {year}
  {2008})}\BibitemShut {NoStop}%
\bibitem [{\citenamefont {Dumm}\ \emph {et~al.}(2009)\citenamefont {Dumm},
  \citenamefont {Faltermeier}, \citenamefont {Drichko}, \citenamefont
  {Dressel}, \citenamefont {M{\'{e}}zi{\`{e}}re},\ and\ \citenamefont
  {Batail}}]{Dumm2009}%
  \BibitemOpen
  \bibfield  {author} {\bibinfo {author} {\bibfnamefont {M.}~\bibnamefont
  {Dumm}}, \bibinfo {author} {\bibfnamefont {D.}~\bibnamefont {Faltermeier}},
  \bibinfo {author} {\bibfnamefont {N.}~\bibnamefont {Drichko}}, \bibinfo
  {author} {\bibfnamefont {M.}~\bibnamefont {Dressel}}, \bibinfo {author}
  {\bibfnamefont {C.}~\bibnamefont {M{\'{e}}zi{\`{e}}re}}, \ and\ \bibinfo
  {author} {\bibfnamefont {P.}~\bibnamefont {Batail}},\ }\href
  {https://link.aps.org/doi/10.1103/PhysRevB.79.195106} {\bibfield  {journal}
  {\bibinfo  {journal} {Phys. Rev. B}\ }\textbf {\bibinfo {volume} {79}},\
  \bibinfo {pages} {195106} (\bibinfo {year} {2009})}\BibitemShut {NoStop}%
\bibitem [{\citenamefont {Aebischer}\ \emph {et~al.}(2001)\citenamefont
  {Aebischer}, \citenamefont {Baeriswyl},\ and\ \citenamefont
  {Noack}}]{Aebischer2001}%
  \BibitemOpen
  \bibfield  {author} {\bibinfo {author} {\bibfnamefont {C.}~\bibnamefont
  {Aebischer}}, \bibinfo {author} {\bibfnamefont {D.}~\bibnamefont
  {Baeriswyl}}, \ and\ \bibinfo {author} {\bibfnamefont {R.~M.}\ \bibnamefont
  {Noack}},\ }\href {\doibase 10.1103/PhysRevLett.86.468} {\bibfield  {journal}
  {\bibinfo  {journal} {Phys. Rev. Lett.}\ }\textbf {\bibinfo {volume} {86}},\
  \bibinfo {pages} {468} (\bibinfo {year} {2001})}\BibitemShut {NoStop}%
\bibitem [{\citenamefont {van Dijk}\ \emph {et~al.}(1986)\citenamefont {van
  Dijk}, \citenamefont {Casteleijn}, \citenamefont {Joosten},\ and\
  \citenamefont {Levine}}]{vanDijk1986}%
  \BibitemOpen
  \bibfield  {author} {\bibinfo {author} {\bibfnamefont {M.~A.}\ \bibnamefont
  {van Dijk}}, \bibinfo {author} {\bibfnamefont {G.}~\bibnamefont
  {Casteleijn}}, \bibinfo {author} {\bibfnamefont {J.~G.~H.}\ \bibnamefont
  {Joosten}}, \ and\ \bibinfo {author} {\bibfnamefont {Y.~K.}\ \bibnamefont
  {Levine}},\ }\href {\doibase 10.1063/1.451588} {\bibfield  {journal}
  {\bibinfo  {journal} {J. Chem. Phys.}\ }\textbf {\bibinfo {volume} {85}},\
  \bibinfo {pages} {626} (\bibinfo {year} {1986})}\BibitemShut {NoStop}%
\bibitem [{\citenamefont {Clarkson}\ and\ \citenamefont
  {Smedley}(1988)}]{Clarkson1988a}%
  \BibitemOpen
  \bibfield  {author} {\bibinfo {author} {\bibfnamefont {M.~T.}\ \bibnamefont
  {Clarkson}}\ and\ \bibinfo {author} {\bibfnamefont {S.~I.}\ \bibnamefont
  {Smedley}},\ }\href {\doibase 10.1103/PhysRevA.37.2070} {\bibfield  {journal}
  {\bibinfo  {journal} {Phys. Rev. A}\ }\textbf {\bibinfo {volume} {37}},\
  \bibinfo {pages} {2070} (\bibinfo {year} {1988})}\BibitemShut {NoStop}%
\bibitem [{\citenamefont {Clarkson}(1988)}]{Clarkson1988b}%
  \BibitemOpen
  \bibfield  {author} {\bibinfo {author} {\bibfnamefont {M.~T.}\ \bibnamefont
  {Clarkson}},\ }\href {\doibase 10.1103/PhysRevA.37.2079} {\bibfield
  {journal} {\bibinfo  {journal} {Phys. Rev. A}\ }\textbf {\bibinfo {volume}
  {37}},\ \bibinfo {pages} {2079} (\bibinfo {year} {1988})}\BibitemShut
  {NoStop}%
\bibitem [{\citenamefont {Pecharrom{\'a}n}\ and\ \citenamefont
  {Moya}(2000)}]{Pecharroman2000}%
  \BibitemOpen
  \bibfield  {author} {\bibinfo {author} {\bibfnamefont {C.}~\bibnamefont
  {Pecharrom{\'a}n}}\ and\ \bibinfo {author} {\bibfnamefont {J.~S.}\
  \bibnamefont {Moya}},\ }\href {\doibase
  10.1002/(SICI)1521-4095(200002)12:4<294::AID-ADMA294>3.0.CO;2-D} {\bibfield
  {journal} {\bibinfo  {journal} {Adv. Mater.}\ }\textbf {\bibinfo {volume}
  {12}},\ \bibinfo {pages} {294} (\bibinfo {year} {2000})}\BibitemShut
  {NoStop}%
\bibitem [{\citenamefont {Nan}\ \emph {et~al.}(2010)\citenamefont {Nan},
  \citenamefont {Shen},\ and\ \citenamefont {Ma}}]{Nan2010}%
  \BibitemOpen
  \bibfield  {author} {\bibinfo {author} {\bibfnamefont {C.-W.}\ \bibnamefont
  {Nan}}, \bibinfo {author} {\bibfnamefont {Y.}~\bibnamefont {Shen}}, \ and\
  \bibinfo {author} {\bibfnamefont {J.}~\bibnamefont {Ma}},\ }\href {\doibase
  10.1146/annurev-matsci-070909-104529} {\bibfield  {journal} {\bibinfo
  {journal} {Ann. Rev. Mater. Res.}\ }\textbf {\bibinfo {volume} {40}},\
  \bibinfo {pages} {131} (\bibinfo {year} {2010})}\BibitemShut {NoStop}%
\bibitem [{\citenamefont {H{\"{o}}vel}\ \emph {et~al.}(2010)\citenamefont
  {H{\"{o}}vel}, \citenamefont {Gompf},\ and\ \citenamefont
  {Dressel}}]{Hovel2010}%
  \BibitemOpen
  \bibfield  {author} {\bibinfo {author} {\bibfnamefont {M.}~\bibnamefont
  {H{\"{o}}vel}}, \bibinfo {author} {\bibfnamefont {B.}~\bibnamefont {Gompf}},
  \ and\ \bibinfo {author} {\bibfnamefont {M.}~\bibnamefont {Dressel}},\ }\href
  {https://link.aps.org/doi/10.1103/PhysRevB.81.035402} {\bibfield  {journal}
  {\bibinfo  {journal} {Phys. Rev. B}\ }\textbf {\bibinfo {volume} {81}},\
  \bibinfo {pages} {35402} (\bibinfo {year} {2010})}\BibitemShut {NoStop}%
\bibitem [{\citenamefont {Dubrov}\ \emph {et~al.}(1976)\citenamefont {Dubrov},
  \citenamefont {Levinshte\v{\i}n},\ and\ \citenamefont {Shur}}]{Dubrov1976}%
  \BibitemOpen
  \bibfield  {author} {\bibinfo {author} {\bibfnamefont {V.~E.}\ \bibnamefont
  {Dubrov}}, \bibinfo {author} {\bibfnamefont {M.~E.}\ \bibnamefont
  {Levinshte\v{\i}n}}, \ and\ \bibinfo {author} {\bibfnamefont {M.~S.}\
  \bibnamefont {Shur}},\ }\href@noop {} {\bibfield  {journal} {\bibinfo
  {journal} {Sov. Phys. JETP}\ }\textbf {\bibinfo {volume} {43}},\ \bibinfo
  {pages} {1050} (\bibinfo {year} {1976})}\BibitemShut {NoStop}%
\bibitem [{\citenamefont {Efros}\ and\ \citenamefont
  {Shklovskii}(1976)}]{Efros1976}%
  \BibitemOpen
  \bibfield  {author} {\bibinfo {author} {\bibfnamefont {A.~L.}\ \bibnamefont
  {Efros}}\ and\ \bibinfo {author} {\bibfnamefont {B.~I.}\ \bibnamefont
  {Shklovskii}},\ }\href@noop {} {\bibfield  {journal} {\bibinfo  {journal}
  {phys. stat. sol. (b)}\ }\textbf {\bibinfo {volume} {76}},\ \bibinfo {pages}
  {475} (\bibinfo {year} {1976})}\BibitemShut {NoStop}%
\bibitem [{\citenamefont {Bergman}\ and\ \citenamefont
  {Imry}(1977)}]{Bergman1977}%
  \BibitemOpen
  \bibfield  {author} {\bibinfo {author} {\bibfnamefont {D.~J.}\ \bibnamefont
  {Bergman}}\ and\ \bibinfo {author} {\bibfnamefont {Y.}~\bibnamefont {Imry}},\
  }\href {\doibase 10.1103/PhysRevLett.39.1222} {\bibfield  {journal} {\bibinfo
   {journal} {Phys. Rev. Lett.}\ }\textbf {\bibinfo {volume} {39}},\ \bibinfo
  {pages} {1222} (\bibinfo {year} {1977})}\BibitemShut {NoStop}%
\bibitem [{\citenamefont {Bergman}(1978)}]{Bergman1978}%
  \BibitemOpen
  \bibfield  {author} {\bibinfo {author} {\bibfnamefont {D.~J.}\ \bibnamefont
  {Bergman}},\ }\href {\doibase https://doi.org/10.1016/0370-1573(78)90009-1}
  {\bibfield  {journal} {\bibinfo  {journal} {Phys. Rep.}\ }\textbf {\bibinfo
  {volume} {43}},\ \bibinfo {pages} {377 } (\bibinfo {year}
  {1978})}\BibitemShut {NoStop}%
\bibitem [{\citenamefont {Choy}(2015)}]{Choy2015}%
  \BibitemOpen
  \bibfield  {author} {\bibinfo {author} {\bibfnamefont {T.~C.}\ \bibnamefont
  {Choy}},\ }\href@noop {} {\emph {\bibinfo {title} {Effective Medium
  Theory}}},\ \bibinfo {edition} {2nd}\ ed.\ (\bibinfo  {publisher} {Oxford
  University Press},\ \bibinfo {address} {Oxford},\ \bibinfo {year}
  {2015})\BibitemShut {NoStop}%
\bibitem [{\citenamefont {H{\'{e}}bert}\ \emph {et~al.}(2015)\citenamefont
  {H{\'{e}}bert}, \citenamefont {S{\'{e}}mon},\ and\ \citenamefont
  {Tremblay}}]{Hebert2015}%
  \BibitemOpen
  \bibfield  {author} {\bibinfo {author} {\bibfnamefont {C.-D.}\ \bibnamefont
  {H{\'{e}}bert}}, \bibinfo {author} {\bibfnamefont {P.}~\bibnamefont
  {S{\'{e}}mon}}, \ and\ \bibinfo {author} {\bibfnamefont {A.-M.~S.}\
  \bibnamefont {Tremblay}},\ }\href {\doibase 10.1103/PhysRevB.92.195112}
  {\bibfield  {journal} {\bibinfo  {journal} {Phys. Rev. B}\ }\textbf {\bibinfo
  {volume} {92}},\ \bibinfo {pages} {195112} (\bibinfo {year}
  {2015})}\BibitemShut {NoStop}%
\bibitem [{\citenamefont {Gati}\ \emph
  {et~al.}(2018{\natexlab{a}})\citenamefont {Gati}, \citenamefont {Winter},
  \citenamefont {Schlueter}, \citenamefont {Schubert}, \citenamefont
  {M{\"{u}}ller},\ and\ \citenamefont {Lang}}]{Gati2018}%
  \BibitemOpen
  \bibfield  {author} {\bibinfo {author} {\bibfnamefont {E.}~\bibnamefont
  {Gati}}, \bibinfo {author} {\bibfnamefont {S.~M.}\ \bibnamefont {Winter}},
  \bibinfo {author} {\bibfnamefont {J.~A.}\ \bibnamefont {Schlueter}}, \bibinfo
  {author} {\bibfnamefont {H.}~\bibnamefont {Schubert}}, \bibinfo {author}
  {\bibfnamefont {J.}~\bibnamefont {M{\"{u}}ller}}, \ and\ \bibinfo {author}
  {\bibfnamefont {M.}~\bibnamefont {Lang}},\ }\href {\doibase
  10.1103/PhysRevB.97.075115} {\bibfield  {journal} {\bibinfo  {journal} {Phys.
  Rev. B}\ }\textbf {\bibinfo {volume} {97}},\ \bibinfo {pages} {75115}
  (\bibinfo {year} {2018}{\natexlab{a}})}\BibitemShut {NoStop}%
\bibitem [{\citenamefont {Urai}\ \emph {et~al.}(2019)\citenamefont {Urai},
  \citenamefont {Furukawa}, \citenamefont {Seki}, \citenamefont {Miyagawa},
  \citenamefont {Sasaki}, \citenamefont {Taniguchi},\ and\ \citenamefont
  {Kanoda}}]{Urai2019}%
  \BibitemOpen
  \bibfield  {author} {\bibinfo {author} {\bibfnamefont {M.}~\bibnamefont
  {Urai}}, \bibinfo {author} {\bibfnamefont {T.}~\bibnamefont {Furukawa}},
  \bibinfo {author} {\bibfnamefont {Y.}~\bibnamefont {Seki}}, \bibinfo {author}
  {\bibfnamefont {K.}~\bibnamefont {Miyagawa}}, \bibinfo {author}
  {\bibfnamefont {T.}~\bibnamefont {Sasaki}}, \bibinfo {author} {\bibfnamefont
  {H.}~\bibnamefont {Taniguchi}}, \ and\ \bibinfo {author} {\bibfnamefont
  {K.}~\bibnamefont {Kanoda}},\ }\href {\doibase 10.1103/PhysRevB.99.245139}
  {\bibfield  {journal} {\bibinfo  {journal} {Phys. Rev. B}\ }\textbf {\bibinfo
  {volume} {99}},\ \bibinfo {pages} {245139} (\bibinfo {year}
  {2019})}\BibitemShut {NoStop}%
\bibitem [{\citenamefont {Gati}\ \emph
  {et~al.}(2018{\natexlab{b}})\citenamefont {Gati}, \citenamefont {Fischer},
  \citenamefont {Lunkenheimer}, \citenamefont {Zielke}, \citenamefont
  {K{\"{o}}hler}, \citenamefont {Kolb}, \citenamefont {von Nidda},
  \citenamefont {Winter}, \citenamefont {Schubert}, \citenamefont {Schlueter},
  \citenamefont {Jeschke}, \citenamefont {Valent{\'{i}}},\ and\ \citenamefont
  {Lang}}]{Gati2018a}%
  \BibitemOpen
  \bibfield  {author} {\bibinfo {author} {\bibfnamefont {E.}~\bibnamefont
  {Gati}}, \bibinfo {author} {\bibfnamefont {J.~K.~H.}\ \bibnamefont
  {Fischer}}, \bibinfo {author} {\bibfnamefont {P.}~\bibnamefont
  {Lunkenheimer}}, \bibinfo {author} {\bibfnamefont {D.}~\bibnamefont
  {Zielke}}, \bibinfo {author} {\bibfnamefont {S.}~\bibnamefont
  {K{\"{o}}hler}}, \bibinfo {author} {\bibfnamefont {F.}~\bibnamefont {Kolb}},
  \bibinfo {author} {\bibfnamefont {H.-A.~K.}\ \bibnamefont {von Nidda}},
  \bibinfo {author} {\bibfnamefont {S.~M.}\ \bibnamefont {Winter}}, \bibinfo
  {author} {\bibfnamefont {H.}~\bibnamefont {Schubert}}, \bibinfo {author}
  {\bibfnamefont {J.~A.}\ \bibnamefont {Schlueter}}, \bibinfo {author}
  {\bibfnamefont {H.~O.}\ \bibnamefont {Jeschke}}, \bibinfo {author}
  {\bibfnamefont {R.}~\bibnamefont {Valent{\'{i}}}}, \ and\ \bibinfo {author}
  {\bibfnamefont {M.}~\bibnamefont {Lang}},\ }\href {\doibase
  10.1103/PhysRevLett.120.247601} {\bibfield  {journal} {\bibinfo  {journal}
  {Phys. Rev. Lett.}\ }\textbf {\bibinfo {volume} {120}},\ \bibinfo {pages}
  {247601} (\bibinfo {year} {2018}{\natexlab{b}})}\BibitemShut {NoStop}%
\bibitem [{\citenamefont {Lunkenheimer}\ \emph {et~al.}(2012)\citenamefont
  {Lunkenheimer}, \citenamefont {M{\"{u}}ller}, \citenamefont {Krohns},
  \citenamefont {Schrettle}, \citenamefont {Loidl}, \citenamefont {Hartmann},
  \citenamefont {Rommel}, \citenamefont {de~Souza}, \citenamefont {Hotta},
  \citenamefont {Schlueter},\ and\ \citenamefont {Lang}}]{Lunkenheimer2012}%
  \BibitemOpen
  \bibfield  {author} {\bibinfo {author} {\bibfnamefont {P.}~\bibnamefont
  {Lunkenheimer}}, \bibinfo {author} {\bibfnamefont {J.}~\bibnamefont
  {M{\"{u}}ller}}, \bibinfo {author} {\bibfnamefont {S.}~\bibnamefont
  {Krohns}}, \bibinfo {author} {\bibfnamefont {F.}~\bibnamefont {Schrettle}},
  \bibinfo {author} {\bibfnamefont {A.}~\bibnamefont {Loidl}}, \bibinfo
  {author} {\bibfnamefont {B.}~\bibnamefont {Hartmann}}, \bibinfo {author}
  {\bibfnamefont {R.}~\bibnamefont {Rommel}}, \bibinfo {author} {\bibfnamefont
  {M.}~\bibnamefont {de~Souza}}, \bibinfo {author} {\bibfnamefont
  {C.}~\bibnamefont {Hotta}}, \bibinfo {author} {\bibfnamefont {J.~A.}\
  \bibnamefont {Schlueter}}, \ and\ \bibinfo {author} {\bibfnamefont
  {M.}~\bibnamefont {Lang}},\ }\href {http://dx.doi.org/10.1038/nmat3400
  http://10.0.4.14/nmat3400
  https://www.nature.com/articles/nmat3400{\#}supplementary-information}
  {\bibfield  {journal} {\bibinfo  {journal} {Nat. Mater.}\ }\textbf {\bibinfo
  {volume} {11}},\ \bibinfo {pages} {755} (\bibinfo {year} {2012})}\BibitemShut
  {NoStop}%
\bibitem [{\citenamefont {Hassan}\ \emph {et~al.}()\citenamefont {Hassan},
  \citenamefont {Thirunavukkuarasu}, \citenamefont {Lu}, \citenamefont
  {Smirnov}, \citenamefont {Zhilyaeva}, \citenamefont {Torunova}, \citenamefont
  {Lyubovskaya},\ and\ \citenamefont {Drichko}}]{Hassan2019}%
  \BibitemOpen
  \bibfield  {author} {\bibinfo {author} {\bibfnamefont {N.~M.}\ \bibnamefont
  {Hassan}}, \bibinfo {author} {\bibfnamefont {K.}~\bibnamefont
  {Thirunavukkuarasu}}, \bibinfo {author} {\bibfnamefont {Z.}~\bibnamefont
  {Lu}}, \bibinfo {author} {\bibfnamefont {D.}~\bibnamefont {Smirnov}},
  \bibinfo {author} {\bibfnamefont {E.}~\bibnamefont {Zhilyaeva}}, \bibinfo
  {author} {\bibfnamefont {S.}~\bibnamefont {Torunova}}, \bibinfo {author}
  {\bibfnamefont {R.}~\bibnamefont {Lyubovskaya}}, \ and\ \bibinfo {author}
  {\bibfnamefont {N.}~\bibnamefont {Drichko}},\ }\href@noop {} {\ }\Eprint
  {http://arxiv.org/abs/arXiv:1905.12740} {arXiv:1905.12740} \BibitemShut
  {NoStop}%
\bibitem [{Note4()}]{Note4}%
  \BibitemOpen
  \bibinfo {note} {While $\epsilon _1$ initially increases upon cooling in
  $\kappa $-(BEDT-TTF)$_2$Cu[N(CN)$_2$]Cl, it peaks at the antiferromagnetic
  transition and reduces at lower $T$. This could be a consequence of the
  metallic fraction first increasing as the insulator-metal phase boundary
  approaches the ambient-pressure position, but then reducing below
  $T_{\protect \rm N}$ because of the negative slope of the boundary between
  antiferromagnet and metal which moves the IMT further away from
  $p=0$.}\BibitemShut {Stop}%
\bibitem [{\citenamefont {Senthil}(2008)}]{Senthil2008}%
  \BibitemOpen
  \bibfield  {author} {\bibinfo {author} {\bibfnamefont {T.}~\bibnamefont
  {Senthil}},\ }\href {https://link.aps.org/doi/10.1103/PhysRevB.78.045109}
  {\bibfield  {journal} {\bibinfo  {journal} {Phys. Rev. B}\ }\textbf {\bibinfo
  {volume} {78}},\ \bibinfo {pages} {45109} (\bibinfo {year}
  {2008})}\BibitemShut {NoStop}%
\bibitem [{\citenamefont {Lee}\ \emph {et~al.}(2016)\citenamefont {Lee},
  \citenamefont {Florens},\ and\ \citenamefont
  {Dobrosavljevi{\'{c}}}}]{Lee2016}%
  \BibitemOpen
  \bibfield  {author} {\bibinfo {author} {\bibfnamefont {T.-H.}\ \bibnamefont
  {Lee}}, \bibinfo {author} {\bibfnamefont {S.}~\bibnamefont {Florens}}, \ and\
  \bibinfo {author} {\bibfnamefont {V.}~\bibnamefont {Dobrosavljevi{\'{c}}}},\
  }\href {https://link.aps.org/doi/10.1103/PhysRevLett.117.136601} {\bibfield
  {journal} {\bibinfo  {journal} {Phys. Rev. Lett.}\ }\textbf {\bibinfo
  {volume} {117}},\ \bibinfo {pages} {136601} (\bibinfo {year}
  {2016})}\BibitemShut {NoStop}%
\bibitem [{\citenamefont {Pustogow}\ \emph
  {et~al.}(2018{\natexlab{c}})\citenamefont {Pustogow}, \citenamefont {Saito},
  \citenamefont {Zhukova}, \citenamefont {Gorshunov}, \citenamefont {Kato},
  \citenamefont {Lee}, \citenamefont {Fratini}, \citenamefont
  {Dobrosavljevi{\'{c}}},\ and\ \citenamefont {Dressel}}]{Pustogow2018spinons}%
  \BibitemOpen
  \bibfield  {author} {\bibinfo {author} {\bibfnamefont {A.}~\bibnamefont
  {Pustogow}}, \bibinfo {author} {\bibfnamefont {Y.}~\bibnamefont {Saito}},
  \bibinfo {author} {\bibfnamefont {E.}~\bibnamefont {Zhukova}}, \bibinfo
  {author} {\bibfnamefont {B.}~\bibnamefont {Gorshunov}}, \bibinfo {author}
  {\bibfnamefont {R.}~\bibnamefont {Kato}}, \bibinfo {author} {\bibfnamefont
  {T.-H.}\ \bibnamefont {Lee}}, \bibinfo {author} {\bibfnamefont
  {S.}~\bibnamefont {Fratini}}, \bibinfo {author} {\bibfnamefont
  {V.}~\bibnamefont {Dobrosavljevi{\'{c}}}}, \ and\ \bibinfo {author}
  {\bibfnamefont {M.}~\bibnamefont {Dressel}},\ }\href {\doibase
  10.1103/PhysRevLett.121.056402} {\bibfield  {journal} {\bibinfo  {journal}
  {Phys. Rev. Lett.}\ }\textbf {\bibinfo {volume} {121}},\ \bibinfo {pages}
  {056402} (\bibinfo {year} {2018}{\natexlab{c}})}\BibitemShut {NoStop}%
\end{thebibliography}

%

\end{document}